\documentclass[prd,onecolumn]{revtex4}
\usepackage{amssymb}
\usepackage{epsfig}
\usepackage{amsmath}
\usepackage{verbatim}

\begin{document}

\title{Chaotic thermalization in Yang-Mills-Higgs theory on a spacial lattice%
}
\author{Ricardo Fariello$^{1}$ and Hilmar Forkel$^{2}$ }
\affiliation{$^{1}$Instituto de F\'{\i}sica Te\'{o}rica, Universidade Estadual Paulista,
01405-900 S\~{a}o Paulo, SP, Brazil}
\affiliation{$^{2}$Institut f\"{u}r Physik, Humboldt-Universit\"{a}t zu Berlin, D-12489
Berlin, Germany }

\begin{abstract}
We analyze the Hamiltonian time evolution of classical SU$\left( 2\right) $
Yang-Mills-Higgs theory with a fundamental Higgs doublet on a spacial
lattice. In particular, we study energy transfer and equilibration processes
among the gauge and Higgs sectors, calculate the maximal Lyapunov exponents
under randomized initial conditions in the weak-coupling regime, where one
expects them to be related to the high-temperature plasmon damping rate, and
investigate their energy and coupling dependence. We further examine
finite-time and finite-size errors, study the impact of the Higgs fields on
the instability of constant non-Abelian magnetic fields, and comment on the
implications of our results for the thermalization properties of hot gauge
fields in the presence of matter.
\end{abstract}

\pacs{?}
\maketitle
\preprint{HU-EP-08-xx, IFT-P.xxx/2008}

\section{Introduction}

A variety of essential physical processes, ranging from ultrarelativistic
heavy-ion collisions \cite{hei04} to the reheating period and phase
transitions in the early Universe \cite{bas06}, proceed at least initially
far from thermodynamic equilibrium and involve abundantly many\
nonperturbative degrees of freedom. The first-principle based theoretical
treatment of such phenomena, which require a quantum field theoretic
description but are inaccessible to Euclidean lattice simulations, is as a
rule beyond present capabilities. Important exceptions to this rule arise,
however, if the underlying amplitudes receive dominant contributions from
classical fields. The latter may be provided, in particular, by bosonic
long-wavelength modes at high temperature $T$ and with energies $E\ll T$
since the Bose-Einstein distribution supplies them with the large occupation
numbers needed to ensure (semi-) classical behavior. In non-Abelian gauge
theories, observables governed by such classical modes are typically of $%
O\left( g^{2}T\right) $ in the weak-coupling regime (where $g$ is the gauge
coupling and $g^{2}T$ sets an inverse classical length scale) and have a
finite classical limit. Prominent examples include the transport
coefficients which control magnetic screening \cite{lin80} and color
diffusion \cite{sel93}, and in particular the static plasmon damping rate 
\cite{bra90}. The latter has direct impact on the local energy and momentum
equilibration processes among hot gauge-field quanta, which were found to
occur over surprisingly short times of less than 1 fm/$c$ in the excited
matter created by ultrarelativistic nuclear collisions at RHIC \cite%
{hei04,pis08}.

Essentially classical nonequilibrium observables of the above type may
therefore be calculated by relating them to real-time\ evolution properties
of classical long-wavelength gauge fields and by simulating those
nonperturbatively on a spacial lattice \cite{mul92,bir294}. Along these
lines, the plasmon damping rate was argued to be proportional to the
classical gluon damping rate and, at least at weak coupling, further to the
maximal\ Lyapunov exponent (MLE) which governs the exponential separation
rate between initially neighboring\ random gauge-field configurations \cite%
{mul92,bir95}. The underlying reasoning is based on the expected ergodicity
of the classical field trajectories and on the relation between exponential
growth and damping rates provided by time reversal symmetry \cite{bir95}. In
the weak-coupling region, furthermore, these relations can be tested
quantitatively by comparison with results from partially resummed thermal
perturbation theory or alternatively from kinetic theory \cite{bra90,bla02}.

Following up on the above arguments, the present paper will deal with the
real-time evolution of classical SU$\left( 2\right) $ Yang-Mills-Higgs (YMH)
theory on spacial lattices of various sizes. A particular focus will be on
the role of the scalar and hence classically treatable matter fields,
provided by the fundamental Higgs doublet, in the chaotic dynamics. The
center piece of the analysis is a systematic survey of the energy and
coupling dependence of a set of maximal Lyapunov exponents designed to cover
representative parts of the weakly coupled YMH phase space. Since our theory
corresponds to the electroweak sector of the standard model with vanishing
Weinberg angle, the resulting MLEs contain information which may be useful
for understanding cosmological nonequilibrium processes during semiclassical
evolution phases of the early Universe, including topological\ structure
formation, baryogenesis \cite{pro97} and potentially cosmic string evolution 
\cite{hin95}.

Moreover, our results will be relevant for the analysis of local
equilibration processes in the highly excited matter produced at the RHIC 
\cite{hei04,pis08} and soon the CERN LHC \cite{arm08} colliders. Indeed, the
chaoticity of the gauge dynamics provides a natural mechanism for entropy
production by soft fields (and the accompanying particle production in the
quantum case), and its most unstable field modes contribute dominantly to
equilibration processes. In particular, our results will give rise to new
estimates for the energy and coupling dependence of the gauge-field damping
rate in the presence of scalar matter. Furthermore, the MLEs should receive
contributions from the non-Abelian plasma instabilities which were recently
argued to accelerate the isotropization and thermalization processes in the
aftermath of high-energy nuclear collisions \cite{mro06}. The underlying
unstable modes could in principle be isolated by numerical techniques
similar to ours. As in chaotic inflation scenarios, furthermore, such
instabilities typically generate nonperturbatively large occupation numbers,
which may extend the reliability of our classical treatment to larger
couplings and lower temperatures. Some of our qualitative results may even
be robust enough to provide guidance on the impact of fundamental quark
fields.

Although our main focus will be on the evolution of random fields, we also
study the impact of the Higgs fields on the instability of a constant
non-Abelian magnetic field. The employed techniques may later be applied to
more complex coherent fields, including classical solutions of YMH theory 
\cite{amb91} and gauge-invariant coherent soft modes \cite{for07}. Studies
of this type could provide new insights into the corresponding quantum
theories. Applied to multimonopole configurations of YMH theory with an
adjoint Higgs field, whose chaotic interactions we have recently studied 
\cite{far05}, they may for example help to clarify the role of chaotic
monopole ensembles in disordering the gauge-theory vacuum.

The paper is organized as follows: in Sec. \ref{ymhlat} we summarize the
formulation of SU$\left( 2\right) $ YMH theory on a Hamiltonian lattice,
derive the corresponding field equations and discuss suitable distance
measures on the space of gauge and Higgs field configurations. Section \ref%
{quant} outlines the main ingredients of our numerical analysis, examines
finite-time and finite-size effects, discusses the time evolution of the
energy transfer between\ the various field sectors, and evaluates the rate
of divergence between initially neighboring random field configurations at
intermediate times. On this basis, we generate in Sec. \ref{lyaetal} a
representative set of maximal Lyapunov exponents, discuss their energy
dependence and relation to the plasmon damping rate, then extend the
analysis by calculating a set of long-time Lyapunov histories, and finally
evaluate the impact of the Higgs fields on the Savvidy instability of
constant non-Abelian magnetic fields. Sec. \ref{impl} puts our results into
context by discussing related nonequlibrium processes in the early Universe
and in the aftermath of high-energy nuclear collisions, and Sec. \ref{sum}
summarizes our main findings and provides some conclusions.

\section{Yang-Mills-Higgs dynamics on a spacial lattice}

\label{ymhlat}

In order to identify and measure chaotic properties of a dynamical system,
one has to follow the evolution of its dynamical variables over sufficiently
long periods of time. A numerical treatment of field theories further
requires to approximate space by a discrete lattice. The analogous handling
of the time variable (as typically implemented in Euclidean spacetime
subject to periodic boundary conditions) is unsuitable for chaos
investigations, however, since it would unacceptably restrict the accessible
evolution times. Hence we resort to the Hamiltonian formulation of lattice
field theory \cite{kog75} in Minkowski space where gauge fields are
restricted to temporal gauge and time remains an unbounded and (in
principle) continuous variable. A further benefit of this formulation is
that residual\ gauge symmetries enforced by Gauss' law can be accurately
preserved during time evolution. In the following subsections we briefly
summarize this approach as it applies to YMH theory and define the distance
measures needed to determine the Lyapunov exponents. (More details can be
found e.g. in Refs. \cite{kog75,chi85}.)

\subsection{Hamiltonian lattice setup}

In the following section we outline pertinent aspects of the Hamiltonian
formulation of 3+1 dimensional SU$\left( 2\right) $ Yang-Mills-Higgs theory
on a spacial cubic lattice subject to periodic boundary conditions. Since
the Higgs field $\phi $ is taken to transform in the fundamental
representation of the gauge group, this theory is equivalent to\ the
electroweak sector of the standard model in the limit of vanishing Weinberg
angle. The gauge is fixed to $A_{0}^{a}=0$, i.e. to Weyl gauge. The unbroken
phase corresponding to a gauge-matter plasma is selected by positive Higgs
mass and interaction terms, which allows for comparison of the results with
(hard-thermal-loop resummed) perturbative results at high temperature below.

The corresponding YMH Hamiltonian can thus be written as%
\begin{eqnarray}
\lefteqn{H=a^{3}\sum_{x,i}\tfrac{1}{2}E_{x,i}^{a}E_{x,i}^{a}+\frac{4}{g^{2}a}%
\sum_{x}\sum_{1\leq i<j\leq 3}\!\left( 1-\tfrac{1}{2}trU_{x,ij}\right) } 
\notag \\
&&+\,a^{3}\sum_{x}\tfrac{1}{2}tr\left( \Dot{\phi}{}_{x}^{\dagger }\Dot{\phi}{%
}_{x}\right) -a\sum_{x,i}tr\left( \phi _{x}^{\dagger }U_{x,i}\phi
_{x+i}\right)   \notag \\
&&+a\sum_{x}\tfrac{1}{2}tr\left( \phi _{x}^{\dagger }\phi _{x}\right) \left[
6+a^{2}\kappa \tfrac{1}{2}tr\left( \phi _{x}^{\dagger }\phi _{x}\right) %
\right]   \label{ham}
\end{eqnarray}%
where $g$ is the gauge coupling, $\kappa $ the Higgs self-coupling, $a$ the
lattice spacing and dots denote time derivatives. The non-Abelian magnetic
field is described by the spacial plaquette%
\begin{equation}
U_{x,ij}\equiv U_{x,i\,}U_{x+i,j\,}U_{x+j,i\,}^{-1}U_{x,j}^{-1}\overset{%
a\rightarrow 0}{\longrightarrow }\exp \left( -iga^{2}F_{x,ij}+O\left(
a^{3}\right) \right)   \label{eqn:up}
\end{equation}%
(with $i,j\in \left\{ 1,2,3\right\} $ and $i\neq j$), i.e. by the ordered
minimal-circumference loop constructed from the link variables 
\begin{equation}
U_{x,i}=\exp \left( -igaA_{x,i}\right)   \label{eqn:link}
\end{equation}%
where $A_{x,i}=A_{x,i}^{a}t^{a}$ is the gauge field and $\sigma ^{a}=2t^{a}$
with $a\in \left\{ 1,2,3\right\} $ are the Pauli matrices. The $U_{x,i}$ are
defined on the link which connects the site $x$ with its neighbor in the
positive $i$ direction. Hence the spacial plaquettes contain the non-Abelian
magnetic field strength components $F_{x,ij}=\frac{1}{a}\left(
A_{x+i,j}-A_{x,j}-A_{x+j,i}+A_{x,i}\right) -ig\left[ A_{x,i},A_{x,j}\right] $
while their electric counterparts $E_{x,i}^{a}=-\dot{A}_{x,i}^{a}$ are
independent variables.\ The first term of the Hamiltonian (\ref{ham})
therefore describes the energy residing in the electric fields while the
second term, 
\begin{equation}
H_{\text{mag}}=\frac{4}{g^{2}a}\sum_{x}\sum_{1\leq i<j\leq 3}\!\left( 1-%
\tfrac{1}{2}trU_{x,ij}\right) \overset{a\rightarrow 0}{\longrightarrow }%
a^{3}\sum_{x,i,j}\tfrac{1}{4}F_{x,ij}^{a}F_{x,ij}^{a}+O\left( a^{4}\right) ,
\label{hmag}
\end{equation}%
approaches the magnetic or potential energy of the gauge field in the naive\
continuum limit.

For the numerical implementation of the SU$\left( 2\right) $ link variables
we have adopted the quaternion representation%
\begin{equation}
U=u^{0}-i\vec{u}\cdot \vec{\sigma}=\left( 
\begin{tabular}{ll}
$u^{0}-iu^{3},$ & $-u^{2}-iu^{1}$ \\ 
$u^{2}-iu^{1},$ & $u^{0}+iu^{3}$%
\end{tabular}%
\right)  \label{quat}
\end{equation}%
(the indices $x,i$ are suppressed) whose real components $u^{\mu }=\left(
u^{0},\vec{u}\right) \in R,$ $\mu \in \left\{ 0,1,2,3\right\} $ satisfy the
constraint $\det U=u^{0}u^{0}+u^{a}u^{a}=1$ and thereby ensure unitarity $%
UU^{\dagger }=1$ as well. The $u^{\mu }$ are thus (four dimensional,
cartesian) coordinates on the SU$\left( 2\right) $ group manifold $S^{3}$.
The representation (\ref{quat}) leads to simple field equations (cf. Sec. %
\ref{eom}) and requires the minimal number of floating point operations to
calculate the product $UV=u^{0}v^{0}-u^{a}v^{a}-i\sigma ^{a}\left(
u^{0}v^{a}+v^{0}u^{a}+\varepsilon ^{abc}u^{b}v^{c}\right) $ of two link
variables. In order to state the initial conditions for the time evolution
of the gauge field, however, we prefer the alternative representation of the
link variable as a rotation of angle $\omega _{\text{G}}$ around the
direction $\hat{n}\left( \vartheta ,\varphi \right) $, i.e.%
\begin{equation}
U=\exp \left( -igA^{a}\frac{\sigma ^{a}}{2}\right) =\cos \left( \frac{\omega
_{\text{G}}}{2}\right) -i\hat{n}\cdot \vec{\sigma}\sin \left( \frac{\omega _{%
\text{G}}}{2}\right)  \label{polar}
\end{equation}%
(suppressing again the indices). In terms of the polar angles $0\leq \omega
_{\text{G}}\leq 2\pi $, $0\leq \vartheta _{\text{G}}\leq \pi $ and $0\leq
\varphi _{\text{G}}\leq 2\pi $ one then has $gA^{a}=\omega _{\text{G}}\hat{n}%
^{a}$ with $\hat{n}^{a}=\left( \sin \vartheta _{\text{G}}\cos \varphi _{%
\text{G}},\sin \vartheta _{\text{G}}\sin \varphi _{\text{G}},\cos \vartheta
_{\text{G}}\right) $ and $u^{0}=\cos \left( \omega _{\text{G}}/2\right) $, $%
u^{a}=\hat{n}^{a}\sin \left( \omega _{\text{G}}/2\right) $. The Higgs field $%
\phi _{x}$ in the fundamental representation of the gauge group is written
in an analogous quaternion representation,%
\begin{equation}
\phi =\phi ^{0}-i\vec{\phi}\cdot \vec{\sigma}=R\left[ \cos \left( \frac{%
\omega _{\text{H}}}{2}\right) -i\hat{n}\cdot \vec{\sigma}\sin \left( \frac{%
\omega _{\text{H}}}{2}\right) \right] ,  \label{h}
\end{equation}%
where the polar decomposition again turns out to be more suitable for
stating the initial conditions (cf. Sec. \ref{init}). In contrast to the
unitary link variables $U$, however, the (square) modulus%
\begin{equation}
R^{2}=\frac{1}{2}tr\left( \phi ^{\dagger }\phi \right)
\end{equation}%
of the Higgs field remains unconstrained.

Exploiting its (classical) scaling properties, the YMH Hamiltonian (\ref{ham}%
) can be reexpressed in terms of the dimensionless variables $\Bar{H}=g^{2}aH
$, $\Bar{E}{}_{x,i}^{a}=ga^{2}E_{x,i}^{a}$, $\Bar{\phi}{}_{x}=ga\phi _{x}$, $%
\bar{\kappa}=\kappa /g^{2}$ and $\bar{t}=t/a$ as%
\begin{equation}
\Bar{H}=\sum_{x}\left[ \varepsilon _{\text{G,el}}\left( x\right)
+\varepsilon _{\text{G,mag}}\left( x\right) +\varepsilon _{\text{H,kin}%
}\left( x\right) +\varepsilon _{\text{H,pot}}\left( x\right) +\varepsilon _{%
\text{G-H}}\left( x\right) \right]   \label{ham2}
\end{equation}%
with the dimensionless energy densities%
\begin{eqnarray}
\varepsilon _{\text{G,el}}\left( x\right)  &=&\sum_{i}\tfrac{1}{2}\Bar{E}{}%
_{x,i}^{a}\Bar{E}{}_{x,i}^{a},\text{ \ \ \ \ }\varepsilon _{\text{G,mag}%
}\left( x\right) =4\sum_{1\leq i<j\leq 3}\!\left( 1-\frac{1}{2}%
trU_{x,ij}\right) , \\
\varepsilon _{\text{H,kin}}\left( x\right)  &=&\frac{1}{2}tr\left( \Dot{\Bar{%
\phi}}{}_{x}^{\dagger }\Dot{\Bar{\phi}}{}_{x}\right) ,\text{ \ \ \ \ \ \ }%
\varepsilon _{\text{G-H}}\left( x\right) =-\sum_{i}tr\left( \Bar{\phi}{}%
_{x}^{\dagger }U_{x,i}\Bar{\phi}{}_{x+i}\right) , \\
\varepsilon _{\text{H,pot}}\left( x\right)  &=&3tr\left( \Bar{\phi}{}%
_{x}^{\dagger }\Bar{\phi}{}_{x}\right) +\frac{\bar{\kappa}}{4}\left[
tr\left( \Bar{\phi}{}_{x}^{\dagger }\Bar{\phi}{}_{x}\right) \right] ^{2},
\label{ehp}
\end{eqnarray}%
where the fields are now functions of $\bar{t}$ and dots represent $d/d\bar{t%
}$. The above form of the Hamiltonian renders the dependence on the total
energy $\Bar{H}$ and the Higgs self-coupling $\bar{\kappa}$, i.e. the two
physical parameters of the YMH system, explicit (whereas the lattice spacing 
$a$ and the gauge coupling $g$ are absorbed into the dimensionless variables
and fields).

\subsection{Field equations}

\label{eom}

The YMH\ Hamiltonian (\ref{ham2}) generates the classical time evolution of
electric, magnetic and Higgs fields. This becomes explicit in the
corresponding first-order Hamilton equations which we derive with the help
of the Poisson brackets%
\begin{equation}
\left\{ X,\Bar{H}\right\} \equiv \frac{\partial X}{\partial q_{s}}\,\frac{%
\partial \Bar{H}}{\partial p_{s}}-\frac{\partial X}{\partial p_{s}}\,\frac{%
\partial \Bar{H}}{\partial q_{s}}  \label{eqn:Pbdef}
\end{equation}%
of the dynamical variables $X$ with the Hamiltonian $\Bar{H}$ (where $%
q_{s},p_{s}$ are the canonically conjugate variables and summation over $s$
is implied). According to the canonical formalism, the time dependence of $X$
is then determined by its Hamilton equation%
\begin{equation}
\Dot{X}=\frac{1}{g^{2}}\left\{ X,\Bar{H}\right\} .  \label{eqn:eqmPbf}
\end{equation}

Specializing Eq. \eqref{eqn:eqmPbf} to the link variable $U_{x,i}$ and
abbreviating $l\equiv \left\{ x,i\right\} $ leads with%
\begin{equation}
\left\{ \Bar{E}{}_{l}^{a},U_{m}\right\} =-ig^{2}t^{a}U_{m}\delta _{lm}
\label{eqn:EUPb}
\end{equation}%
to the equation of motion%
\begin{equation}
\Dot{U}{}_{l}=\frac{1}{g^{2}}\left\{ U_{l},\Bar{H}\right\} =i\Bar{E}{}%
_{l}U_{l}\quad  \label{eqn:emU}
\end{equation}%
where $\Bar{E}{}_{l}=\Bar{E}{}_{l}^{a}t^{a}$. In the quaternion
representation (\ref{quat}) this equation reads%
\begin{equation}
\Dot{u}{}_{l}^{0}=\frac{1}{2}\Bar{E}{}_{l}^{a}u_{l}^{a},~\ \ \ \ \ \Dot{u}{}%
_{l}^{a}=-\frac{1}{2}\left( \Bar{E}{}_{l}^{a}u_{l}^{0}+\epsilon ^{abc}\Bar{E%
}{}_{l}^{b}u_{l}^{c}\right)  \label{eqn:emq}
\end{equation}%
and maintains, in particular, the time-independence of the unitarity
constraint, i.e. 
\begin{equation}
\Dot{u}{}_{l}^{0}u_{l}^{0}+\Dot{u}{}_{l}^{a}u_{l}^{a}=0.  \label{eqn:Unorm}
\end{equation}%
Hamilton's equation for the non-Abelian electric field strengths $\Bar{E}{}%
_{x,i}^{a}$, which are the canonically conjugate momenta of the link
variables, similarly becomes 
\begin{equation}
\Dot{\Bar{E}}{}_{x,i}^{a}=\frac{1}{g^{2}}\left\{ \bar{E}{}_{x,i}^{a},\Bar{H}%
\right\} =\frac{i}{2}\sum_{j}tr\left[ \sigma ^{a}\left(
U_{x,ij}-U_{x,ij}^{-1}\right) \right] +\frac{i}{2}tr\left( \Bar{\phi}{}%
_{x}^{\dagger }\sigma ^{a}U_{x,i}\Bar{\phi}{}_{x+i}\right)  \label{emE}
\end{equation}%
where the sum goes over the four plaquettes which contain the link $\left\{
x,i\right\} $.

The Hamilton equations for the Higgs field, its canonical momentum $\pi $
and their Hermitian conjugates are analogously found to be

\begin{eqnarray}
\Dot{\Bar{\phi}}_{x} &=&\frac{1}{g^{2}}\left\{ \Bar{\phi}_{x},\bar{H}%
\right\} =\frac{g}{a}\pi _{x}^{\dagger }=\tfrac{1}{2}tr\left( \Dot{\Bar{\phi}%
}_{x}\right)  \label{h1} \\
\Dot{\Bar{\phi}}{}_{x}^{\dagger } &=&\frac{1}{g^{2}}\left\{ \Bar{\phi}%
_{x}^{\dagger },\bar{H}\right\} =\frac{g}{a}\pi _{x}  \label{h2}
\end{eqnarray}%
and%
\begin{eqnarray}
\Dot{\pi}_{x} &=&\frac{1}{g^{2}}\left\{ \pi _{x},\bar{H}\right\} =-\frac{a}{%
2g}tr\left\{ \left[ 6+\bar{\kappa}tr\left( \Bar{\phi}{}_{x}^{\dagger }\Bar{%
\phi}{}_{x}\right) \right] \Bar{\phi}{}_{x}^{\dagger }-2\sum_{i}\Bar{\phi}{}%
_{x-i}^{\dagger }U_{x-i,i}\right\} ,  \label{h3} \\
\Dot{\pi}_{x}^{\dagger } &=&\frac{1}{g^{2}}\left\{ \pi _{x}^{\dagger },\bar{H%
}\right\} =-\frac{a}{2g}tr\left\{ \left[ 6+\bar{\kappa}tr\left( \Bar{\phi}{}%
_{x}^{\dagger }\Bar{\phi}{}_{x}\right) \right] \Bar{\phi}{}%
_{x}-2\sum_{i}U_{x,i}\Bar{\phi}{}_{x+i}\right\} .  \label{h4}
\end{eqnarray}%
In order to prepare for an efficient numerical solution of this system, we
rewrite it in terms of two second-order equations,

\begin{eqnarray}
\Ddot{\Bar{\phi}}{}_{x} &=&-\bar{\kappa}tr\left( \Bar{\phi}{}_{x}^{\dagger }%
\Bar{\phi}{}_{x}\right) \Bar{\phi}{}_{x}-6\,\Bar{\phi}{}_{x}+2\sum_{i}U_{x,i}%
\Bar{\phi}{}_{x+i},  \label{eqn:emppd} \\
\Ddot{\Bar{\phi}}{}_{x}^{\dagger } &=&-\bar{\kappa}tr\left( \Bar{\phi}{}%
_{x}^{\dagger }\Bar{\phi}{}_{x}\right) \Bar{\phi}{}_{x}^{\dagger }-6\,\Bar{%
\phi}{}_{x}^{\dagger }+2\sum_{i}\Bar{\phi}{}_{x-i}^{\dagger }U_{x-i,i},
\label{eqn:emhcppd}
\end{eqnarray}%
and then combine those, by adding the Hermitian conjugate of Eq. (\ref%
{eqn:emhcppd}) to Eq. (\ref{eqn:emppd}), into 
\begin{equation}
\Ddot{\Bar{\phi}}{}_{x}=-\bar{\kappa}tr\left( \Bar{\phi}{}_{x}^{\dagger }%
\Bar{\phi}{}_{x}\right) \Bar{\phi}{}_{x}-6\,\Bar{\phi}{}_{x}+\sum_{i}\left(
\,U_{x,i}\Bar{\phi}{}_{x+i}+U_{x-i,i}^{-1}\Bar{\phi}{}_{x-i}\right) .
\label{eqn:emphi}
\end{equation}

Finally, we recall that the full YMH dynamics in temporal gauge is only
recovered after supplementing Hamilton's equations (\ref{eqn:emq}), (\ref%
{emE}) and (\ref{eqn:emphi}) by Gauss' law%
\begin{equation}
\sum_{i=1}^{3}\left[ \bar{E}_{x,i}^{a}-\frac{1}{2}tr\left(
U_{x,-i}^{-1}\sigma ^{a}U_{x,-i}\sigma ^{b}\right) \bar{E}_{x-i,i}^{b}\right]
=\bar{\rho}_{x}^{a}  \label{Gss}
\end{equation}%
which acts as a constraint. Since its Poisson bracket with the Hamiltonian (%
\ref{ham2}) vanishes, Gauss' law is preserved under time evolution. Above,
we have defined the dimensionless non-Abelian charge density 
\begin{equation}
\bar{\rho}_{x}^{a}=\frac{1}{2}tr\left[ \text{Im}\left( \dot{\bar{\phi}}%
_{x}^{\dagger }\bar{\phi}_{x}\sigma ^{4-a}\right) \right]  \label{j}
\end{equation}%
carried by the Higgs field.

\subsection{Distance measures for gauge and Higgs field configurations}

\label{dist}

The chaotic behavior of dynamical systems reveals itself in an exponential
sensitivity of their time evolution to small changes in the initial
conditions. The quantitative characterization of this sensitivity requires a
distance measure on the field configuration space (i.e. a metric). More
specifically, in the YMH system one has to monitor the separation between a
reference configuration $(U_{l},\phi )$ and its neighbor $(U_{l}^{\prime
},\phi ^{\prime })=(U_{l}+\delta U,\phi +\delta \phi )$. We will use
individual distance measures in the gauge and Higgs sectors for this
purpose, in order to determine the distance growth rate between two
initially nearby gauge and Higgs field configurations individually.

In the gauge sector, we adopt the gauge-invariant metric \cite{mul92} 
\begin{equation}
d_{\text{G}}[U_{l},U_{l}^{\prime }]=\frac{1}{2N_{\text{p}}}%
\sum_{p}\,\left\vert trU_{p}-trU_{p}^{\prime }\right\vert   \label{du}
\end{equation}%
(where $N_{\text{p}}=3N^{3}$ is the total number of plaquettes on a lattice
with $N$ sites per spacial dimension) to measure the distance between
gauge-field configurations. In the continuum limit the distance measured by
the metric (\ref{du}) becomes proportional to the difference between the
potential energies of reference and neighboring gauge fields. In the Higgs
sector we employ the metric \cite{bir94} 
\begin{equation}
d_{\text{H}}[\phi ,\phi ^{\prime }]=\frac{1}{N^{3}}\sum_{x}\,\left\vert
\left( R_{x}\right) ^{2}-\left( R_{x}^{\prime }\right) ^{2}\right\vert 
\label{dh}
\end{equation}%
which is gauge invariant as well.

Since the lattice gauge group, and consequently the $3\left(
N_{c}^{2}-1\right) N^{3}$ dimensional space of magnetic SU$\left(
N_{c}\right) $ gauge-field configurations on a lattice with $N$ sites per
dimension, is compact and of nontrivial topology, more and more field
configurations approach the same distance $d_{\text{G}}$ when $N$ increases.
For the same reason, the distance (\ref{du}) is bounded from above, and for
fixed total energy an analogous bound applies to the whole phase space.%
\textbf{\ }These bounds lead to an eventual saturation of the distance
growth. Although this does not limit the principal effectiveness of the
measures $d_{\text{G,H}}$ for determining the Lyapunov exponents (see
below), it adds to the typical \textquotedblleft
finite-time\textquotedblright\ uncertainties encountered in their numerical
analysis. Other sources of finite-time errors arise from the need to
extrapolate the numerical results to the $t\rightarrow \infty $ limit in
which the MLEs are formally defined, and for $N\rightarrow \infty $ from the
exponentially growing distances between chaotic trajectories which
eventually overburden the floating point number representation capacities of
any computer.

The standard approach for keeping finite-time errors of MLEs under control
is to periodically rescale the distances \cite{ben80} after time intervals $%
\tau $ and to extrapolate the numerical results for $\ln d_{\text{G,H}%
}\left( t\right) /t$ to infinite times. This approach has been used to
calculate several MLEs in non-Abelian gauge theories \cite%
{bir94,gon94,mue96,hei97} and to determine the whole Lyapunov spectrum on
small lattices \cite{gon94}. We have adopted the same technique for the
calculation of several long-time trajectories to be discussed in Secs. \ref%
{edistr} and \ref{ltevol}. In these cases, we found it advantageous to
employ the alternative distances measure%
\begin{equation}
d_{\text{G}}^{\left( \text{alt}\right) }[U_{l},\bar{E}_{l};U_{l}^{\prime },%
\bar{E}_{l}^{\prime }]=\left\{ \sum_{l}\left[ \left( \bar{E}_{l}-\bar{E}%
_{l}^{\prime }\right) ^{2}+\left( U_{l}-U_{l}^{\prime }\right) ^{2}\right]
\right\} ^{1/2}  \label{du2}
\end{equation}%
in the phase space of the gauge fields, which is a variant of the measure
used in Ref. \cite{hei97}, and 
\begin{equation}
d_{\text{H}}^{\left( \text{alt}\right) }[\bar{\phi},\bar{\phi}^{\prime
}]=\left\{ \sum_{x}\sum_{\alpha =0}^{3}\left[ \left( \dot{\bar{\phi}}%
_{x}^{\alpha }-\dot{\bar{\phi}}_{x}^{\alpha \prime }\right) ^{2}+\left( \bar{%
\phi}_{x}^{\alpha }-\bar{\phi}_{x}^{\alpha \prime }\right) ^{2}\right]
\right\} ^{1/2},  \label{dh2}
\end{equation}
adopted from Ref. \cite{hei97}, in the Higgs sector. Of course, the
resulting Lyapunov exponents should not depend on the choice of distance
measure. We have checked this for several examples and confirmed that the
deviations between the MLE values obtained from the metrics (\ref{du}), (\ref%
{dh}) and (\ref{du2}), (\ref{dh2}) indeed remain well below the one-percent
level.

Nevertheless, even under rescaling the practically achievable evolution
times remain limited by the available computer resources. In fact, even in
pure SU$\left( 2\right) $ Yang-Mills theory \cite{mue96} systematic
extrapolation errors turned out to become negligible only after evolution
times of the order of $10^{5}$ lattice units. To make matters worse, we will
find below that the equilibration between gauge and Higgs fields proceeds at
a far slower pace than among the gauge fields alone (cf. Sec. \ref{edistr}),
and that as a consequence substantially longer evolution times are required
to suppress such extrapolation errors in YMH theory. Adherence to one of our
main goals, namely to calculate a rather exhaustive set of MLEs in the
weak-coupling parameter and phase space, will therefore require a
compromise. Indeed, to cover the relevant initial parameter space (on
lattices of several different sizes) requires the calculation of $O\left(
10^{2}\right) $ trajectory pairs and thus forces us to limit the individual
evolution times.

Fortunately, size and systematics of finite-time errors can be estimated on
the basis of the long-time energy balance (cf. Sec. \ref{edistr}) and a few
long-time orbits (cf. Sec. \ref{ltevol}). Since we are mainly interested in
the systematic energy-, coupling- and lattice-size dependence of the MLEs
(rather than in their precise numerical values), furthermore, the competing
goals of error suppression and calculability can be reconciled reasonably
well. Our compromise will be to follow the majority of our distance
histories only until they have saturated (without rescaling), which yields
sufficiently accurate MLE estimates for most of our purposes. Rescaling will
be used, on the other hand, for the long-time trajectories which we need to
examine the energy transfer and equilibration processes between the gauge
and Higgs field sectors in Sec. \ref{edistr}, and for the analysis of the
MLE's finite-time errors and saturation properties in Sec. \ref{ltevol}.

\section{Field initialization, energy balance and distance evolution}

\label{quant}

In the following section we discuss in turn the initialization of the
neighboring field configurations, the distribution of the total energy over
the different field sectors, and the time evolution of the distance between
initially adjacent field configurations.

\subsection{Initial conditions}

\label{init}

Our first task will be to generate a representative set of phase-space
trajectories for pairs of specifically initialized reference\ YMH fields $%
(U_{l},\phi )$ and their neighboring configurations $(U_{l}^{\prime },\phi
^{\prime })=(U_{l}+\delta U,\phi +\delta \phi )$ at sufficiently small
distances $d_{\text{G}}[U_{l},U_{l}^{\prime }]$ and $d_{\text{H}}[\phi ,\phi
^{\prime }]$. The resulting distance evolution histories will provide one of
the foundations for our subsequent analysis of the maximal Lyapunov
exponents. In the present subsection, we select a set of 77 initial
conditions for the reference trajectories such that the weak-coupling region
of the YMH phase space is covered with sufficient resolution. In order to
allow for direct comparison with a previously calculated MLE, we follow the
initialization procedure of Ref. \cite{hei97}. The resulting sample of
field-pair trajectories will be considerably larger than that of preceding
MLE calculations in gauge theories and include results from substantially
larger lattice volumes (with up to 30$^{3}$ sites).

\begin{figure}[tbp]
\begin{center}
\includegraphics[height = 9cm]{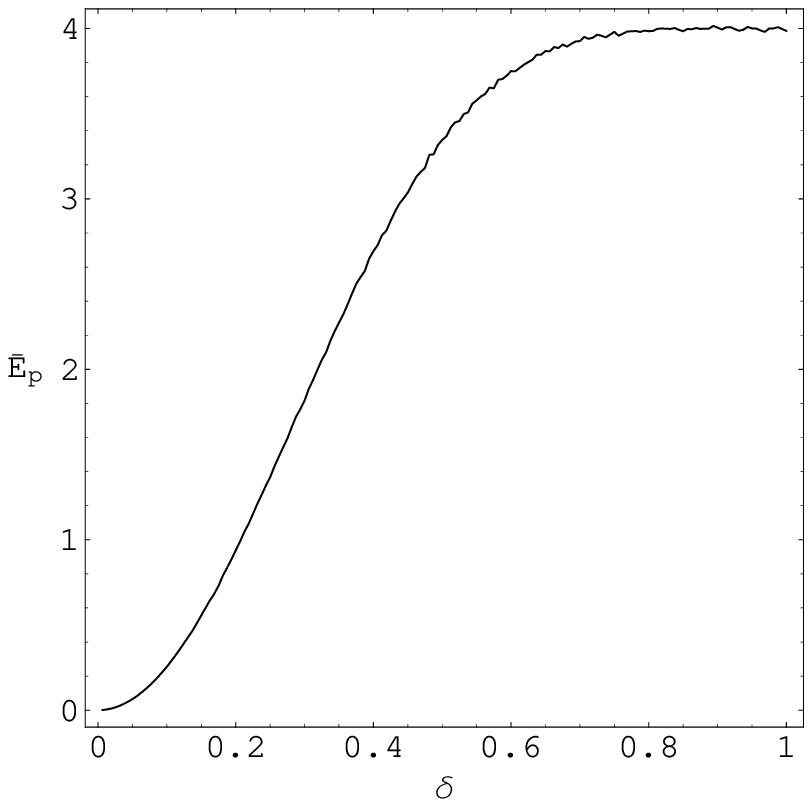}
\end{center}
\caption{The average energy per plaquette $\bar{E}_{\text{p}}$ as a function
of the initialization parameter $\protect\delta $.}
\label{epdel}
\end{figure}

In order to satisfy Gauss' law (\ref{Gss}) initially (and consequently over
the whole time evolution), we set the non-Abelian electric field and the
time derivative of the Higgs field at the initial time $\bar{t}_{0}=0$ equal
to zero, i.e. 
\begin{equation}
\Bar{E}{}_{x,i}^{a}\left( 0\right) =0,\text{ \ \ \ \ }\Dot{\Bar{\phi}}{}%
_{x}\left( 0\right) =0,
\end{equation}%
which implies $\bar{\rho}_{x}^{a}\left( 0\right) =0$ (cf. Eq. (\ref{j})).
Hence the initial kinetic energies of all fields vanish while the potential
energies are finite and ensure that the system starts far from equilibrium.
The link variables $U_{l}$ are initialized by randomly choosing the isospin
directions $\vartheta _{\text{G}},\varphi _{\text{G}}$ of the gauge
potential $A_{l}^{a}=\omega _{\text{G,}l}\Hat{n}{}_{l}^{a}(\vartheta _{\text{%
G},l},\varphi _{\text{G},l})$ from their full domains, while the initial
value of the amplitude $\omega _{\text{G}}$ is chosen randomly over the
restricted domain $\omega _{\text{G,}l}\in \lbrack 0,2\pi \delta ]$ with $%
\delta \leq 1$. The value of the parameter $\delta $ therefore controls the
average gauge-field energy per plaquette, $\bar{E}_{\text{p}}\left( \delta
\right) $, which grows as $\delta ^{2}$ for $\delta \ll 1$ and saturates in
the limit $\delta \rightarrow 1$ at the value $\bar{E}_{\text{p}}=4$ (cf.
Fig. \ref{epdel}). The upper bound on $\bar{E}_{\text{p}}$ arises from the
fact that the magnetic part (\ref{hmag}) of the Hamiltonian (\ref{ham}) is
uniformly bounded by the SU$\left( 2\right) $ group volume, $\bar{H}_{\text{%
mag}}=ag^{2}H_{\text{mag}}\leq 24N^{3}$. The Higgs field (\ref{h}), finally,
is initialized by choosing its angular variables $\omega _{\text{H}}$, $%
\vartheta _{\text{H}}$ and $\varphi _{\text{H}}$ randomly from their full
domains while keeping the dimensionless amplitude $\bar{R}_{x}\equiv gaR_{x}$
fixed at the same value $\bar{R}$ for all $\vec{x}$. As a consequence, the
initial (potential) energy of the Higgs field is determined by the amplitude 
$\bar{R}$ and the coupling $\bar{\kappa}$.

The above initialization scheme characterizes any phase-space trajectory on
a given lattice by three parameters $\delta $, $\bar{R}$ and $\bar{\kappa}$
which determine the average initial energy of both the gauge and Higgs
fields. In addition, we will vary the lattice size, specified by the number $%
N$ of sites per dimension, so that each of our field-pair histories can be
uniquely labeled by a quadruple of values for $\delta ,\bar{R},\bar{\kappa}$
and $N$. Our maximal lattice size with $N=30$ is chosen to substantially
reduce potential finite-size effects of previous studies \cite%
{bir94,hei97,bir96} on considerably smaller lattices. The main benefit of
the random angle initialization is that it equips the initial configurations
with a specific average energy density, or equivalently with a temperature $T
$ which the fields will reach after equilibration. In our context, this is
important because the temperature dependence of the MLEs is used to relate
them to the static plasmon damping rates. Moreover, the resulting MLE values
will turn out to be (within errors) independent of the random part of a
given starting configuration, which indicates that the autocorrelation
functions of the fields have decayed sufficiently strongly before the MLEs
are measured (see below).

In order to stay safely inside the validity range of the semiclassical
approximation, and to be able to relate our findings to perturbative
results, we will restrict our simulations to the weak-coupling regime. As
pointed out in Ref. \cite{hei97}, this requires that the energy contributed
by the Higgs mass term dominates over the Higgs self-interaction energy,
i.e. 
\begin{equation}
\bar{\kappa}\bar{R}^{2}<1,  \label{lambineq}
\end{equation}%
and that the magnetic gauge-field energy dominates over the gauge-Higgs
interaction energy (which implies a weak gauge-Higgs coupling), or 
\begin{equation}
\bar{R}^{2}<\delta   \label{delineq}
\end{equation}%
(for maximal field amplitudes). Both conditions also improve the eventual
equipartition of the electric, magnetic and Higgs field energies because
they prevent the total energy from strongly exceeding the bounded magnetic
energy. The lower bound (\ref{delineq}) on $\delta $ additionally limits
finite-size effects [cf. Eq. (\ref{fslim})]. Furthermore, $\delta $ values
too close to 1 should be avoided in order to keep lattice-spacing artefacts
under control and to remain sufficiently close to the continuum limit (cf.
Eq. (\ref{lslim})).

For each of the initial reference configurations $(U_{l},\phi )$ created
according to the above procedure, we also generate a neighboring
configuration $(U_{l}^{\prime },\phi ^{\prime })$ separated from $%
(U_{l},\phi )$ by distances $d_{\text{G}}(t=0)\lesssim 5\times 10^{-7}\ $and 
$d_{\text{H}}(t=0)\lesssim 10^{-17}$. This is achieved by randomly choosing
slight variations of all the reference configuration's field angles in the
range $\delta \omega _{\text{G}}$, $\delta \vartheta _{\text{G}}$, $\delta
\varphi _{\text{G}}$, $\delta \omega _{\text{H}}$, $\delta \vartheta _{\text{%
H}}$, $\delta \varphi _{\text{H}}\in \left[ -\varepsilon ,\varepsilon \right]
$ where $\varepsilon =10^{-6}$. We then integrate the field equations of
Sec. \ref{eom} for each of these configuration pairs \footnote{%
This involves the time evolution of (12 (link field) + 9 (iso-electric
field) + 8 (Higgs field)) $\times $ $N^{3}$ variables, i.e. altogether of $%
7.83\times 10^{5}$ coupled variables for our largest lattices with $N=30$.}
by means of a fourth-order Runge-Kutta algorithm and determine the time
evolution of the distances $d_{\text{G}}$ and $d_{\text{H}}$. The
integration time step should be much shorter than the lattice spacing $a$,
and is additionally chosen small enough to ensure energy conservation with
an accuracy of more than eight significant digits (after each step). The
maximal violation of the constraints $\det {U}_{l}{=1}$ after a single time
step of length $\Delta t=10^{-4}a$ is then about $10^{-12}$ at each link. In
order to avoid the accumulation of these round-off errors, we further
rescale the link variables after each step such that their determinant
remains exactly unity\ (and Eq. \eqref{eqn:Unorm} exactly satisfied). We
convinced ourselves that Gauss' law (\ref{Gss}) then remains satisfied to
better than five significant digits after each integration step.

\subsection{Energy distribution over gauge and Higgs fields}

\label{edistr}

We now turn to the energy transfer processes between the electric, magnetic
and Higgs fields which contain crucial information on the nonequilibrium
dynamics and quantitative thermalization properties of the YMH system. In
our context, this information will be particularly helpful for
understanding, qualitatively estimating and reducing the finite-time errors
which afflict the calculation of the MLEs, and for putting the relation
between the MLEs and the plasmon damping rates on a more solid footing. For
several long-time trajectories, we have therefore recorded the evolution of
the energies per degree of freedom stored in the electric field, $E_{\text{el%
}}=H_{\text{el}}/\left( 6N^{3}\right) $, in the magnetic gauge field, $E_{%
\text{mag}}=H_{\text{mag}}/\left( 6N^{3}\right) $, and in the Higgs field, $%
E_{\text{H}}=H_{\text{H}}/\left( 4N^{3}\right) ,$ over the unprecedentedly
long time periods $0\leq t\leq 20000a$ . In the following, we will often
express the total energy (per degree of freedom) $E_{\text{G}}$ of the gauge
field in terms of the \emph{average} energy per plaquette $E_{\text{p}}$ as 
\begin{equation}
E_{\text{G}}=E_{\text{el}}+E_{\text{mag}}=\frac{1}{2}E_{\text{p}},
\label{eg}
\end{equation}%
and frequently encounter the total YMH energy per degree of freedom, $%
E=\left( 6E_{\text{G}}+4E_{\text{H}}\right) /10$, as well.

Typical results for the time evolution of the different energies are plotted
in Fig. \ref{ens} with $\delta =0.2$, $\bar{R}=0.2$, $\bar{\kappa}=$ $1$, $%
N=10$ and in Fig. \ref{ens2} with $\delta =1$ and otherwise unchanged
initial values. They confirm and extend the observation of Ref. \cite{hei97}
that the energy equilibration between the electric, magnetic and Higgs field
sectors of YMH theory proceed over two drastically different time scales (at
least in the weak-coupling regime). Indeed, even when initialized in highly
nonequilibrium configurations, as selected in Sec. \ref{init}, the electric
and magnetic gauge sectors can be seen to equilibrate very rapidly, namely
after only a few lattice time units $a$. The Higgs field's potential and
kinetic energies, which are not shown separately in Figs. \ref{ens} and \ref%
{ens2}, equilibrate over an approximately equal relaxation time.\ (Generally
the gauge and Higgs sectors reach different temperatures, however, according
to the amount of energy stored in them by the initial conditions.) In
contrast, the mutual thermalization of gauge and Higgs sectors typically
requires 4 to 5 orders of magnitude more time. In fact, the energy transfer
between the two sectors becomes appreciable only after a few hundred time
units and takes several thousand more to essentially complete for $\delta =1$%
, and many more for $\delta =0.2$. Moreover, for the maximal $\delta =1$
moderate deviations from complete equipartition of the energy remain visible
in Fig. \ref{ens2} even after 10000 time units have elapsed. This may be a
consequence of lattice-spacing artefacts which are maximal at $\delta =1$
(cf. Sec. \ref{init}). The huge discrepancy between the two characteristic
relaxation scales can be largely attributed to the initial conditions of
Sec. \ref{init} which keep the system close to the weak-coupling and
continuum limits.

The gauge-field energy (\ref{eg}) can be directly related to the temperature 
$T$ which the gauge fields reach after times $t\gg \lambda ^{-1}$ (where $%
\lambda $ is the MLE). At sufficiently weak coupling (among the field
oscillators) one has \cite{amb92} 
\begin{equation}
T=\frac{3}{2\left( N_{c}^{2}-1\right) }E_{\text{p}}  \label{tfromep}
\end{equation}%
for the gauge group SU$\left( N_{c}\right) $ and thus $E_{\text{G}}=T$ for $%
N_{c}=2$. This relation will be relevant for the evaluation and
interpretation of the MLEs which we extract in Sec. \ref{lya} from the
distance growth rates after the gauge fields became members of a prethermal
ensemble. As mentioned in the introduction, $l_{\text{cl}}=\left(
g^{2}T\right) ^{-1}=\left( g^{2}E_{\text{G}}\right) ^{-1}$ acts as a
classical length scale in hot quantum gauge-theory amplitudes which depend
(to leading order in thermal perturbation theory) on $g$ and $T$ exclusively
in the combination $g^{2}T$. This observation suggests additional conditions
for keeping lattice artefacts in such amplitudes under control \cite{mue96}.
More specifically, in order to remain sufficiently close to the continuum
limit the lattice spacing $a$ should be much smaller than $l_{\text{cl}}$,
i.e.%
\begin{equation}
\bar{E}_{\text{G}}=ag^{2}E_{\text{G}}\ll 1,  \label{lslim}
\end{equation}%
and in order to avoid finite-size effects the extent $Na$ of the cubic
lattice has to be much larger than $l_{\text{cl}}$, i.e.%
\begin{equation}
N\bar{E}_{\text{G}}\gg 1.  \label{fslim}
\end{equation}%
As expected, these conditions require $N\gg 1$, and the upper bound (\ref%
{lslim}) on $\bar{E}_{\text{G}}$ furthermore ensures that the underlying
lattice structure cannot be resolved by the gauge fields. (Of course, for $%
a\rightarrow 0$ one will eventually encounter UV singularities of
Rayleigh-Jeans type in some amplitudes, signalling the onset of
indispensable quantum corrections to the classical field statistics.)

\subsection{Divergence of neighboring field trajectories in phase space 
\label{div}}

In the following section we analyze the time evolution of the distances $d_{%
\text{G}}$ and $d_{\text{H}}$ between pairs of initially adjacent random
field configurations which were generated according to the procedure of Sec. %
\ref{init} and followed until saturation. The MLEs and their parameter and
in particular energy dependence will then be extracted from the growth rate
of the logarithmic distances in Sec. \ref{lya}. In order to cover the
relevant phase space, we select a representative set of values for the
parameters $\delta $, $\bar{R}$, $\bar{\kappa}$ and $N$ which characterize
any initial configuration. The initial, homogeneous Higgs field amplitude is
fixed at $\bar{R}=0.2$ for all trajectory pairs, which allows for a
quantitative comparison with a configuration studied in Ref. \cite{hei97}.
To stay sufficiently close to the weak-coupling and continuum limits then
requires, according to Eq. (\ref{lambineq}), that the Higgs self-coupling is
bounded by $\bar{\kappa}<\allowbreak 25$, and as a consequence of Eq. (\ref%
{delineq}) that the initial magnetic (and total) gauge-field energy is
restricted by $\delta >0.04$. As mentioned above, the bound on $\delta $
also helps to avoid significant finite-size artefacts [cf. Eq. (\ref{fslim}%
)] and allows to extract the approximate MLEs with reasonable accuracy even
after rather small evolution times (see below).

Guided by the above arguments, we generate trajectory pairs for 11 values of 
$\delta \in \left[ 0.05,1\right] $. For each of them, we plot the resulting $%
\ln d{_{\text{G}}}\left( t\right) $ (black lines)\ and $\ln d{_{\text{H}}}%
\left( t\right) $ (grey lines) in Fig. \ref{fig1} at fixed Higgs
self-coupling $\bar{\kappa}=1$ on lattices of four different sizes
corresponding to $N=6,10,20$ and $30$, and in Fig. \ref{fig2} on a $N=20$
lattice with the Higgs coupling values $\bar{\kappa}=1,8,16$ and $24$. (The
configuration pair studied in Ref. \cite{hei97} on a relatively small
lattice with $N=10$ and $\delta =0.2$, $\bar{R}=0.2,$ $\bar{\kappa}=1$ is
therefore included in our sample.) The corresponding logarithmic distances $%
\ln d_{\text{G,H}}\left( t\right) $ for the 11 $\delta $ values are grouped
into three sets which are separately plotted in panels (a) -- (c) of Figs. %
\ref{fig1} and \ref{fig2}: in panel (a) we display $\ln d_{\text{G,H}}$ for
the five largest values $\delta =1.0,0.5,0.45,0.4,0.35$, in panel (b) for
the values $\delta =0.3,0.25,0.2,0.15$ and in panel (c) for the two smallest
values $\delta =0.1,0.05$. All $\delta $ values except for the smallest
(i.e. $\delta =0.05$, which is most strongly affected by finite-size
artefacts) store more energy in the gauge than in the Higgs sector.

The essential characteristic which all logarithmic distance histories of
Figs. \ref{fig1} and \ref{fig2}\ share is that after a latency period of
varying length they start to rise at least approximately linearly with $t/a$
before reaching a time-independent saturation plateau (which lies somewhat
outside the plotted $\bar{t}$ domain for $\delta =0.05$) at the maximal
distance in the compact phase space. Distance saturation at large $\bar{t}%
=t/a$ is a consequence of the compactness of the lattice gauge group and
could be avoided by periodical rescaling (cf. Sec. \ref{dist}). The linear
regions and the underlying exponential growth rates between initially almost
identical field configurations reveal an exponential sensitivity of the
distance evolution to the initial conditions, i.e. the standard hallmark of
temporal chaos. Not surprisingly, the fields grow apart at a faster pace
when their energy increases, i.e. the slopes in Figs. \ref{fig1} and \ref%
{fig2} grow with $\delta $. For each field trajectory, furthermore, the
linear regions of both $\ln d{_{\text{G}}}\left( t\right) $ and $\ln d{_{%
\text{H}}}\left( t\right) $ have the same average slopes. This result
differs from a previous estimate for one trajectory \cite{bir96} and will be
discussed further in Sec. \ref{lya}. Moreover, for $\delta \lesssim 0.2$ the
latency period, which is hardly noticeable for larger $\delta $, expands and
the linear growth becomes increasingly modulated by oscillations whose
frequency increases with $\delta $. This behavior was observed in YM theory
as well and can be traced to the impact of the next-to-maximal Lyapunov
exponents which grows when the maximal exponent decreases \cite{mul92}.
Obviously, these oscillations reduce the accuracy with which the maximal
Lyapunov exponent can be determined from the slopes of $\ln d{_{\text{G,H}}}%
\left( t\right) $ in the linear regions (see below).

Figures \ref{fig1} and \ref{fig2} further show that for all field
trajectories (except that with $\delta =0.05$) $\ln d{_{\text{H}}}\left(
t\right) $ stays below $\ln d{_{\text{G}}}\left( t\right) $. This reflects
the smaller amount of energy initially stored in the\ Higgs sector for $%
\delta >0.05$ (cf. Sec. \ref{init}) and will change during the long-time
evolution to be discussed in Sec. \ref{ltevol}. In addition, the height of
the saturation plateaus of $\ln d_{\text{G}}\left( t\right) $ decreases
slightly with $\delta $ while that of $\ln d{_{\text{H}}}\left( t\right) $
remains constant. This may indicate that the maximal magnetic gauge-field
distance (\ref{du}) is reached only when sufficient gauge-field energy is
available. Apart from these differences in the saturation behavior, however,
even the modulation patterns of $\ln d_{\text{G}}\left( t\right) $ and $\ln d%
{_{\text{H}}}\left( t\right) $ are very similar. This suggests that, despite
the small gauge-Higgs coupling ensured by Eq. (\ref{delineq}), the time
dependence of the gauge and Higgs components of at least the most unstable
mode has already synchronized after a few lattice time units.

The qualitative dependence of the results on the lattice size, i.e. on $N\in
\left\{ 6,10,20,30\right\} $ with the lattice UV cutoff $a^{-1}$ kept fixed,
can be judged by comparing the distance histories in Fig. \ref{fig1}. Figure %
\ref{fig1} a contains the results for $1\geq \delta \geq 0.35$. Although the
fields are randomly initialized, the curves with identical $\delta $ but
different $N$ clearly cluster, i.e. in accord with the bound (\ref{fslim})
essentially no finite-size effects can be observed in the covered $N$ and $%
\delta $ regions (while lattice-spacing effects should become noticeable for 
$\delta $ close to unity [cf. Eq. (\ref{lslim})]). Indications for a similar 
$N$ independence were found in pure YM theory \cite{mul92}. This may suggest
that the most unstable modes, i.e. those which dominantly drive the chaotic
time evolution of initially adjacent configurations, have for sufficiently
large initial magnetic field energy (corresponding to $\delta \geq 0.35$)
typical wavelengths which are small enough to be accommodated by even the
largest considered IR cutoff (corresponding to $N=6$), or in other words
that these most chaotic modes essentially fit inside a periodic $\left(
6a\right) ^{3}$ lattice volume. As shown in Figs. \ref{fig1}b and \ref{fig1}%
c, however, for smaller $\delta \lesssim 0.2$ finite-size corrections become
visible in the average slopes of the (increasingly oscillation modulated)
linear regions of $\ln d{_{\text{G,H}}}\left( t\right) $ and cause them to
differ more strongly. A systematic trend in the $N$ dependence of these
slopes cannot be discerned in our data, however, whereas in pure YM theory
the slope was found to increase on smaller lattices \cite{mue96}.

Figure \ref{fig2} reveals the qualitative dependence of the distance
histories on the Higgs self-coupling $\bar{\kappa}$. In the range $\bar{%
\kappa}\in \left\{ 1,8,16,24\right\} $ (for $N=20$) it bears several
qualitative similarities with the $N$ dependence of Fig. \ref{fig1}. To
begin with, a $\bar{\kappa}$ dependence is hardly noticeable for large $%
\delta $ while the slope of the linear regions becomes increasingly $\bar{%
\kappa}$ dependent towards smaller values of $\delta $, although again
without a perceivable systematic trend. The, as a whole, only mild
sensitivity of the slopes to $\bar{\kappa}$ is probably a consequence of the
fact that even $\ln d_{\text{H}}\left( t\right) $ is mainly determined by
the most unstable gauge-field fluctuations and hence relatively insensitive
to the self-interactions of the Higgs field. After full equilibration
between the gauge and Higgs sector has taken place, the $\bar{\kappa}$
dependence of the slopes may therefore be systematically enhanced (if
distance saturation is avoided by periodical rescaling), as we will indeed
find in Sec. \ref{ltevol}. Since a more strongly self-coupled Higgs sector
would absorb energy from the gauge sector (in which for $\delta \geq 0.1$
more initial energy is stored, cf. Sec. \ref{init}) faster, it should
similarly increase the $\bar{\kappa}$ dependence of the slopes.

To summarize, all members of the representative set of distance histories
(in the weakly coupled, symmetric YMH phase) discussed above increase
exponentially and thereby exhibit chaotic behavior. For $\delta \gtrsim 0.3$
the $\ln d_{\text{G,H}}\left( t\right) $ become practically independent of
the Higgs coupling $\bar{\kappa}$ and (for $N\geq 6$) of the lattice volume.

\section{Lyapunov exponents, scaling behavior and damping rates}

\label{lyaetal}

In the following section we proceed to the quantitative evaluation of the
maximal Lyapunov exponents for randomized and coherent initial conditions,
and we discuss their energy dependence and relation to the plasmon damping
rates.

\subsection{Maximal Lyapunov exponents of randomly initialized fields}

\label{lya}

The analysis of the last section showed that all our 77 randomly initialized
field pairs belong to the chaotic part of the YMH phase space. This suggests
that in the unbroken phase of YMH theory chaotic behavior is either
universal (i.e. exists for all energies) or at least prevalent in most of
the weakly coupled phase space \footnote{%
Strictly universal chaos is almost impossible to establish numerically. In
the extreme long-wavelength limit the classical Yang-Mills phase space
(without Higgs field), for example, contains tiny regions in which stable,
i.e. nonergodic trajectories exist \cite{zak94,dah90}.}. In order to
quantify this behavior, we will now evaluate the classic measure for the
chaoticity of a dynamical system, i.e. the maximal Lyapunov exponent\ $%
\lambda $ or equivalently the exponential growth rate of the distance
between initially neighboring dynamical variables. We are going to extract
the MLEs from the numerical results of Sec. \ref{dist} by averaging the time
histories $d_{\text{G}}\left( \bar{E},\bar{\kappa},N;t\right) $ and $d_{%
\text{H}}\left( \bar{E},\bar{\kappa},N;t\right) $ of the gauge and Higgs
field distance measures over the time interval $\Delta $ during which they
remain in the linear regime, i.e.%
\begin{equation}
\lambda _{\text{G},\text{H}}\left( \bar{E},\bar{\kappa},N\right)
=\left\langle \frac{d}{dt}\ln \frac{d_{\text{G,H}}\left( \bar{E},\bar{\kappa}%
,N;t\right) }{d_{\text{G,H}}\left( \bar{E},\bar{\kappa},N;0\right) }%
\right\rangle _{\Delta }.
\end{equation}%
Note that we have replaced the dependence on the initialization parameter $%
\delta $ with that on the total (dimensionless) energy $\bar{E}$ of the YMH
system, and that we suppressed the dependence on the remaining
initialization parameter, the Higgs amplitude $\bar{R}$, which is kept at
the same value for all our trajectories (cf. Sec. \ref{init}). We further
recall that the above method for obtaining the MLEs becomes increasingly
error-prone towards lower energies where equilibration proceeds more slowly
while the impact of the next-to-maximal Lyapunov exponents grows and
generates modulations of $\ln d_{\text{G,H}}\left( t\right) $ with
decreasing frequency. Similar problems were encountered in Ref. \cite{mul92}
and will be tamed below by periodically rescaling the distance measures (cf.
Sec. \ref{ltevol}).

Our numerical results for the dimensionless MLEs $\bar{\lambda}_{\text{G},%
\text{H}}\left( \bar{E},\bar{\kappa},N\right) :=a\lambda _{\text{G},\text{H}%
}\left( \bar{E},\bar{\kappa},N\right) $, based on the 77 field-pair
evolution histories of Sec. \ref{div}, are collected in Table 1. 
\begin{table}[tp]
\begin{center}
\begin{tabular}{|c||c|c|c|c|c|c|c|}
\hline
& 
\begin{tabular}{c}
$N=6$ \\ 
$\bar{\kappa}=1$%
\end{tabular}
& 
\begin{tabular}{c}
$N=10$ \\ 
$\bar{\kappa}=1$%
\end{tabular}
& 
\begin{tabular}{c}
$N=20$ \\ 
$\bar{\kappa}=1$%
\end{tabular}
& 
\begin{tabular}{c}
$N=30$ \\ 
$\bar{\kappa}=1$%
\end{tabular}
& 
\begin{tabular}{c}
$N=20$ \\ 
$\bar{\kappa}=8$%
\end{tabular}
& 
\begin{tabular}{c}
$N=20$ \\ 
$\bar{\kappa}=16$%
\end{tabular}
& 
\begin{tabular}{c}
$N=20$ \\ 
$\bar{\kappa}=24$%
\end{tabular}
\\ \hline\hline
$\bar{E}=0.04374$ & 
\begin{tabular}{c}
$0.06507$ \\ 
$0.07521$%
\end{tabular}
& 
\begin{tabular}{c}
$0.07445$ \\ 
$0.07900$%
\end{tabular}
& 
\begin{tabular}{c}
$0.06179$ \\ 
$0.06799$%
\end{tabular}
& 
\begin{tabular}{c}
$0.04563$ \\ 
$0.04713$%
\end{tabular}
& 
\begin{tabular}{c}
$0.06276$ \\ 
$0.06422$%
\end{tabular}
& 
\begin{tabular}{c}
$0.02922$ \\ 
$0.02517$%
\end{tabular}
& 
\begin{tabular}{c}
$0.04862$ \\ 
$0.04712$%
\end{tabular}
\\ \hline
$\bar{E}=0.10076$ & 
\begin{tabular}{c}
$0.07394$ \\ 
$0.07343$%
\end{tabular}
& 
\begin{tabular}{c}
$0.05838$ \\ 
$0.05820$%
\end{tabular}
& 
\begin{tabular}{c}
$0.05052$ \\ 
$0.05004$%
\end{tabular}
& 
\begin{tabular}{c}
$0.05617$ \\ 
$0.05633$%
\end{tabular}
& 
\begin{tabular}{c}
$0.05246$ \\ 
$0.05270$%
\end{tabular}
& 
\begin{tabular}{c}
$0.05981$ \\ 
$0.06028$%
\end{tabular}
& 
\begin{tabular}{c}
$0.03972$ \\ 
$0.03997$%
\end{tabular}
\\ \hline
$\bar{E}=0.19028$ & 
\begin{tabular}{c}
$0.09117$ \\ 
$0.09156$%
\end{tabular}
& 
\begin{tabular}{c}
$0.07269$ \\ 
$0.07320$%
\end{tabular}
& 
\begin{tabular}{c}
$0.09783$ \\ 
$0.09833$%
\end{tabular}
& 
\begin{tabular}{c}
$0.09281$ \\ 
$0.09336$%
\end{tabular}
& 
\begin{tabular}{c}
$0.08522$ \\ 
$0.08561$%
\end{tabular}
& 
\begin{tabular}{c}
$0.09264$ \\ 
$0.09330$%
\end{tabular}
& 
\begin{tabular}{c}
$0.08767$ \\ 
$0.08849$%
\end{tabular}
\\ \hline
$\bar{E}=0.30482$ & 
\begin{tabular}{c}
$0.13357$ \\ 
$0.13349$%
\end{tabular}
& 
\begin{tabular}{c}
$0.13348$ \\ 
$0.13410$%
\end{tabular}
& 
\begin{tabular}{c}
$0.13017$ \\ 
$0.13075$%
\end{tabular}
& 
\begin{tabular}{c}
$0.13751$ \\ 
$0.13803$%
\end{tabular}
& 
\begin{tabular}{c}
$0.13876$ \\ 
$0.13888$%
\end{tabular}
& 
\begin{tabular}{c}
$0.13665$ \\ 
$0.13655$%
\end{tabular}
& 
\begin{tabular}{c}
$0.13435$ \\ 
$0.13394$%
\end{tabular}
\\ \hline
$\bar{E}=0.43527$ & 
\begin{tabular}{c}
$0.19660$ \\ 
$0.19844$%
\end{tabular}
& 
\begin{tabular}{c}
$0.20009$ \\ 
$0.20126$%
\end{tabular}
& 
\begin{tabular}{c}
$0.20985$ \\ 
$0.21099$%
\end{tabular}
& 
\begin{tabular}{c}
$0.20906$ \\ 
$0.21011$%
\end{tabular}
& 
\begin{tabular}{c}
$0.22112$ \\ 
$0.22164$%
\end{tabular}
& 
\begin{tabular}{c}
$0.20315$ \\ 
$0.20313$%
\end{tabular}
& 
\begin{tabular}{c}
$0.20032$ \\ 
$0.19922$%
\end{tabular}
\\ \hline
$\bar{E}=0.57202$ & 
\begin{tabular}{c}
$0.28580$ \\ 
$0.28877$%
\end{tabular}
& 
\begin{tabular}{c}
$0.25783$ \\ 
$0.25994$%
\end{tabular}
& 
\begin{tabular}{c}
$0.29550$ \\ 
$0.29795$%
\end{tabular}
& 
\begin{tabular}{c}
$0.29339$ \\ 
$0.29540$%
\end{tabular}
& 
\begin{tabular}{c}
$0.28875$ \\ 
$0.29021$%
\end{tabular}
& 
\begin{tabular}{c}
$0.28975$ \\ 
$0.29024$%
\end{tabular}
& 
\begin{tabular}{c}
$0.29050$ \\ 
$0.28895$%
\end{tabular}
\\ \hline
$\bar{E}=0.70604$ & 
\begin{tabular}{c}
$0.36801$ \\ 
$0.37857$%
\end{tabular}
& 
\begin{tabular}{c}
$0.39740$ \\ 
$0.39802$%
\end{tabular}
& 
\begin{tabular}{c}
$0.39422$ \\ 
$0.39841$%
\end{tabular}
& 
\begin{tabular}{c}
$0.39359$ \\ 
$0.39638$%
\end{tabular}
& 
\begin{tabular}{c}
$0.39232$ \\ 
$0.39354$%
\end{tabular}
& 
\begin{tabular}{c}
$0.37328$ \\ 
$0.37355$%
\end{tabular}
& 
\begin{tabular}{c}
$0.38446$ \\ 
$0.38460$%
\end{tabular}
\\ \hline
$\bar{E}=0.82974$ & 
\begin{tabular}{c}
$0.46159$ \\ 
$0.46487$%
\end{tabular}
& 
\begin{tabular}{c}
$0.46709$ \\ 
$0.47145$%
\end{tabular}
& 
\begin{tabular}{c}
$0.48083$ \\ 
$0.48596$%
\end{tabular}
& 
\begin{tabular}{c}
$0.46971$ \\ 
$0.47334$%
\end{tabular}
& 
\begin{tabular}{c}
$0.47193$ \\ 
$0.47344$%
\end{tabular}
& 
\begin{tabular}{c}
$0.47730$ \\ 
$0.47800$%
\end{tabular}
& 
\begin{tabular}{c}
$0.48521$ \\ 
$0.48630$%
\end{tabular}
\\ \hline
$\bar{E}=0.93767$ & 
\begin{tabular}{c}
$0.50948$ \\ 
$0.51818$%
\end{tabular}
& 
\begin{tabular}{c}
$0.53934$ \\ 
$0.53885$%
\end{tabular}
& 
\begin{tabular}{c}
$0.53530$ \\ 
$0.53656$%
\end{tabular}
& 
\begin{tabular}{c}
$0.54047$ \\ 
$0.54306$%
\end{tabular}
& 
\begin{tabular}{c}
$0.52785$ \\ 
$0.52760$%
\end{tabular}
& 
\begin{tabular}{c}
$0.53862$ \\ 
$0.53854$%
\end{tabular}
& 
\begin{tabular}{c}
$0.54295$ \\ 
$0.54123$%
\end{tabular}
\\ \hline
$\bar{E}=1.02672$ & 
\begin{tabular}{c}
$0.56526$ \\ 
$0.56125$%
\end{tabular}
& 
\begin{tabular}{c}
$0.56621$ \\ 
$0.56722$%
\end{tabular}
& 
\begin{tabular}{c}
$0.58570$ \\ 
$0.58918$%
\end{tabular}
& 
\begin{tabular}{c}
$0.58284$ \\ 
$0.58643$%
\end{tabular}
& 
\begin{tabular}{c}
$0.57729$ \\ 
$0.57799$%
\end{tabular}
& 
\begin{tabular}{c}
$0.58613$ \\ 
$0.58700$%
\end{tabular}
& 
\begin{tabular}{c}
$0.57540$ \\ 
$0.57636$%
\end{tabular}
\\ \hline
$\bar{E}=1.22634$ & 
\begin{tabular}{c}
$0.63279$ \\ 
$0.64082$%
\end{tabular}
& 
\begin{tabular}{c}
$0.65405$ \\ 
$0.65811$%
\end{tabular}
& 
\begin{tabular}{c}
$0.65635$ \\ 
$0.65997$%
\end{tabular}
& 
\begin{tabular}{c}
$0.65077$ \\ 
$0.65550$%
\end{tabular}
& 
\begin{tabular}{c}
$0.64431$ \\ 
$0.64588$%
\end{tabular}
& 
\begin{tabular}{c}
$0.65041$ \\ 
$0.65104$%
\end{tabular}
& 
\begin{tabular}{c}
$0.64915$ \\ 
$0.65008$%
\end{tabular}
\\ \hline
\end{tabular}%
\end{center}
\caption{Maximal Lyapunov exponents $\bar{\protect\lambda}_{\text{G}}=a%
\protect\lambda _{\text{G}}$ (upper entries) and $\bar{\protect\lambda}_{%
\text{H}}=a\protect\lambda _{\text{H}}$ (lower entries) as a function of
total energy $\bar{E}=g^{2}Ea$, Higgs self-coupling $\bar{\protect\kappa}$
and number $N$ of lattice sites per dimension. }
\label{LyaTab}
\end{table}
A first glance at the table confirms the qualitative trends which we noticed
in our discussion of Figs. \ref{fig1} and \ref{fig2} in Sec. \ref{div}.
Besides the expected increase of the $\bar{\lambda}_{\text{G},\text{H}}$
with $\bar{E}$ (or $\delta $), which we will analyze quantitatively in Sec. %
\ref{edep}, the data show fluctuations in the statistically expected range
of about 10\% for different Higgs self-couplings and lattice sizes, but
except for the smallest $\bar{E}$ no obvious systematic dependence on either 
$\bar{\kappa}$ or $N$. At the considered intermediate times (i.e. after
separate preequilibration of gauge and Higgs sectors but before their mutual
thermalization is complete) and at least at intermediate energies $\bar{E}$
or $\bar{E}_{\text{p}}$ systematic\ finite-size and lattice-spacing effects
are therefore small. Furthermore, the above results indicate that the Higgs
sector plays a rather minor role in the chaoticity of the full YMH system,
at least at the weak couplings which the initial conditions of Sec. \ref%
{init} implement. The most unstable mode, which in large part drives the
chaotic behavior, should therefore be controlled mainly by the gauge
dynamics. As a consequence, reasonable estimates for the MLEs can be
extracted at the preequilibration stage and the MLE values of the SU$\left(
2\right) $ YMH system should be similar to those of pure SU$\left( 2\right) $
YM\ theory \cite{mul92,mue96}, which is confirmed by the results in Table 1.

While we find the Higgs sector to have only limited impact on the chaotic
YMH dynamics in the weakly coupled \emph{symmetric} phase, it may be useful
to recall the results of Refs. \cite{mat81,ber85} in the homogeneous limit,
i.e. for wavelengths much larger than the inverse amplitudes $\left\vert
A\right\vert ^{-1},\left\vert \phi \right\vert ^{-1}$, which reveal a more
dramatic role of the Higgs field in the broken phase (even at nonzero
Weinberg angle). This is a consequence of the dynamically generated
gauge-field mass in the broken phase which is known to damp (and beyond a
critical value to fully suppress) chaotic behavior \cite{bir94}. (We note in
passing that chaos is not only damped by gauge-field masses generated via
spontaneous symmetry breaking, but also by those due to quantum fluctuations
according to the Coleman-Weinberg mechanism \cite{mat97}, topological
excitations, polarization of the heat bath at finite temperature, and
external charges \cite{bir294}.) In Ref. \cite{ber85} chaotic behavior was
observed \footnote{%
in the additional presence of an Abelian gauge field which probably does not
significantly affect the chaotic properties} only beyond the threshold
energy $\bar{E}_{\text{th}}\simeq 0.3$ (showing that chaos is not universal
in the broken phase), and for the energy $\bar{E}=5.07\gg \bar{E}_{\text{th}}
$ the MLE was found to be $\bar{\lambda}\simeq 0.25$ \cite{ber85}, i.e. an
order of magnitude smaller than our value $\bar{\lambda}\simeq 2.75$ in the
unbroken phase (which we linearly extrapolate [cf. Sec. \ref{edep}] from the
values in Table 1 up to $\bar{E}=5.07$). Since constant fields with their
few degrees of freedom can exhibit stronger chaoticity and thus produce
larger MLEs than our randomized initial configurations, this comparison
gives a quantitative idea of how much the chaotic YMH instability is damped
by the Higgs mechanism in the broken phase.

Another issue which can be addressed quantitatively on the basis of the data
in Table 1 is the relation between the maximal Lyapunov exponents $\bar{%
\lambda}_{\text{G}}$ and $\bar{\lambda}_{\text{H}}$, which are obtained from
the gauge and Higgs field distance measures (\ref{du}) and (\ref{dh}),
respectively. This relation was subject to some debate, in particular at
strong coupling \cite{bir96,hei97}. After an exploratory study in Ref. \cite%
{bir94}, Ref. \cite{bir96} provided a first lattice estimate for YMH theory.
The $\bar{\lambda}_{\text{H}}$ extracted from the growth rate of the Higgs
field distance measure was found to become smaller than $\bar{\lambda}_{%
\text{G}}$ when the Higgs self-coupling $\bar{\kappa}$ increases. At $\bar{%
\kappa}=24$ and for $N=10$, in particular, $\bar{\lambda}_{\text{H}}$ was
estimated in Ref. \cite{bir96} to be about 15\% smaller than $\bar{\lambda}_{%
\text{G}}$. Comparison with the static gauge and Higgs boson damping rate in
(resummed) thermal perturbation theory then cast doubt on their relation to
the same $\bar{\lambda}_{\text{G}}$ and led to the speculation that the
Higgs damping rate may instead be related to $\bar{\lambda}_{\text{H}}$ \cite%
{bir96}. These ideas were later questioned in Ref. \cite{hei97} whose
improved calculation found $\bar{\lambda}_{\text{G}}$\ and $\bar{\lambda}_{%
\text{H}}$ to agree, although only for one trajectory pair at fixed energy
and $\delta =0.2,$ $\bar{\kappa}=1,$ $\bar{R}=0.2$ and $N=10$.

In our case, the $\bar{\lambda}_{\text{G}}$ (upper entries)\ and $\bar{%
\lambda}_{\text{H}}$ (lower entries) values in Table 1 agree within errors
(at the percent level) in all of the covered YMH phase space, with the
deviations slightly decreasing for increasing $\bar{E}$ and $\bar{\kappa}$.
Our results therefore show that the finding of Ref. \cite{hei97} was not an
accidental outcome of one specific initialization choice but that indeed%
\begin{equation}
\bar{\lambda}_{\text{G}}\left( \bar{E},\bar{\kappa},N\right) \simeq \bar{%
\lambda}_{\text{H}}\left( \bar{E},\bar{\kappa},N\right) .  \label{lgeqlh}
\end{equation}%
Since the individual relaxation times $\tau $ of the gauge and Higgs sectors
are set by the inverse MLEs, i.e. $\tau _{\text{G,H}}\simeq \lambda _{\text{%
G,H}}^{-1}$, Eq (\ref{lgeqlh}) naturally explains the observation $\tau _{%
\text{G}}\sim \tau _{\text{H}}$ in Sec. \ref{edistr}, i.e. the fact that
both gauge and Higgs sectors (separately) self-thermalize over about the 
\emph{same} relaxation time. Equation (\ref{lgeqlh}) further squares with
the general expectation that the maximally unstable field mode of a
dynamical system, i.e. the mode associated with the MLE, dominates the
exponential distance growth. Hence the MLEs should be independent of the
metric used to extract them (modulo constant factors which depend on the
field powers involved in the definition of the metric). A possible exception
to this rule may arise, however, if the distance measure is blind\ to the
maximally unstable eigenmode. In Ref. \cite{hei97} it was argued that such a
situation occurs in YMH theory at large coupling $\bar{\kappa}$, where the
quartic Higgs self-interaction dominates the potential Higgs energy (\ref%
{ehp}): the amplitude $R_{x}$ then remains practically unchanged during time
evolution and decouples from the maximally unstable gauge-field mode to
which the Higgs distance measure (\ref{dh}) consequently becomes
insensitive. At the still relatively weak coupling $\bar{\kappa}=24$, and
the energy $\bar{E}\sim 0.8$ of Ref. \cite{bir96} where $d_{\text{H}}\left(
t\right) $ is rather strongly time dependent [cf. Fig. \ref{fig2}(a)],
however, our results for $\bar{\lambda}_{\text{H}}$ and $\bar{\lambda}_{%
\text{G}}$ differ by only about 1\%. This suggests that the 15\% deviation
found in Ref. \cite{bir96} should mainly be attributed to numerical
uncertainties.

\subsection{Energy dependence and relation to plasmon damping rate}

\label{edep}

As already mentioned, the dependence of the MLEs on the gauge-field energy
per degree of freedom, $E_{\text{G}}=E_{\text{p}}/2$ (cf. Eq. (\ref{eg})),
and on the total energy $E$ of the YMH system is a particularly important
issue. In pure SU$\left( 2\right) $ \cite{mul92,bir94} and SU$\left(
3\right) $ \cite{bir94,gon93} Yang-Mills theory (whose scaling properties
imply that the dimensionless Lyapunov exponent $\bar{\lambda}=\lambda a$ can
only depend on $\bar{E}_{\text{p}}=g^{2}aE_{\text{p}}$), the approximately
linear relation%
\begin{equation}
\lambda _{N_{c}}\simeq c_{N_{c}}g^{2}E_{\text{p}}  \label{lamErel}
\end{equation}%
with $c_{2}\simeq 0.17$ and $c_{3}\simeq 0.10$ was established numerically
in the weak-coupling regime. (An improved SU$\left( 2\right) $ analysis and
a careful discussion of the involved errors \cite{mue96}, triggered by
questions raised in Ref. \cite{nie96}, later confirmed the results of Refs. 
\cite{mul92,bir94}.) The empirical relation (\ref{lamErel}) helps to clarify
the physical role of the MLEs in hot \emph{quantum} gauge theory. Since the
Lyapunov exponents were extracted at times $t\gg \tau _{\text{G}}\simeq
\lambda ^{-1}$, i.e. after the gauge sector has preequilibrated \footnote{%
This fast thermalization should happen at any level of coarse-graining since
the growth rate of the entropy density increases with the volume \cite{gon94}%
.}, the thermal gauge-field ensemble has according to Eq. (\ref{tfromep})
reached the temperature $T_{\text{G}}=E_{\text{p}}/2$ (at sufficiently large
average plaquette energy $E_{\text{p}}$). Together with Eq. (\ref{lamErel})
this implies the linear relationship $\lambda _{N_{c}}=\tilde{c}_{N_{c}}T_{%
\text{G}}$, and comparison with the static plasmon damping rate $\gamma
_{0,N_{c}}$ of hot quantum SU$\left( N_{c}\right) $ YM theory, as calculated
to leading order in hard-thermal-loop resummed perturbation theory \cite%
{bra90}, then revealed the at first rather unexpected relation \cite%
{mul92,bir94,gon93} 
\begin{equation}
\lambda _{N_{c}}\simeq 2\gamma _{0,N_{c}}  \label{lamgam}
\end{equation}%
for $N_{c}=2,3$. [The factor of 2 arises because the growth rate of the
distance (\ref{du}) is twice that of the distance between the gauge fields.]
Subsequently, Eq. (\ref{lamgam}) has been derived under a few heuristic
assumptions (in particular on the ergodicity of the gauge-field evolution)
in Ref. \cite{bir95}.

On the basis of the rather exhaustive data set in Table 1, we are now able
to address the analogous question of how the MLEs are related to the average
plaquette and total energies in the weak-coupling regime of YMH theory. In
Fig. \ref{MLE1vsEp} we plot the MLEs for $\bar{\kappa}=1$ on lattices with $%
N\in \left\{ 6,10,20,30\right\} $ (corresponding to the first four columns
of Table 1) in the full range of average plaquette energies $0\lesssim
ag^{2}E_{\text{p}}\equiv \bar{E}_{\text{p}}\lesssim 4$. Figure~\ref{MLE2vsEp}
contains all remaining MLEs of Table 1, i.e. those for $\bar{\kappa}=\left\{
1,8,16,24\right\} $ at $N=20$. The straight lines also drawn in Figs. \ref%
{MLE1vsEp} and \ref{MLE2vsEp} are the best linear fits to the data: 
\begin{equation}
\bar{\lambda}_{\text{G,H}}\left( E_{\text{p}}\right) \simeq 0.17ag^{2}E_{%
\text{p}}\simeq \frac{1}{6}\bar{E}_{\text{p}}.  \label{lamEp}
\end{equation}%
The figures show that the MLEs indeed depend within errors linearly on the
average energy $E_{\text{p}}$ per plaquette, as in YM theory. In fact, the
linearity of $\bar{\lambda}\left( \bar{E}_{\text{p}}\right) $ seems to be a
nontrivial consequence of the non-Abelian nature of the gauge group. (The
MLEs of scalar $\phi ^{4}$ theory and Abelian U$\left( 1\right) $ gauge
theory, in contrast, were found to vanish in the continuum limit $%
a\rightarrow 0$ \cite{bir94}.) Remarkably, even the slope of the linear
relation (\ref{lamEp}) is almost identical to that in SU$\left( 2\right) $
Yang-Mills theory \cite{mul92,bir94,mue96}. (It is also consistent with the
value of the ratio $\bar{\lambda}/\bar{E}_{\text{p}}$ which was extracted
from the trajectory with $\delta =0.2,$ $\bar{\kappa}=1,$ $\bar{R}=0.2$ and $%
N=10$ in Ref. \cite{hei97}.)

Equation (\ref{lamEp}) implies that for identical \emph{gauge} field energy
the MLEs of YM and YMH\ theory are approximately equal. This provides our
main evidence for the maximally unstable YMH mode to belong primarily to the
gauge sector, and suggests that the chaoticity and equilibration properties
of the Higgs sector are mediated by this gauge-field mode as well (at least
at weak coupling and if the major part of the initial energy is stored in
the gauge sector). It also makes it more plausible that the MLEs of YMH
theory are related to the \emph{gauge}\ field damping rates \cite%
{bir94,hei97}. Furthermore, it is consistent with the approximately equal
relaxation times $\tau _{\text{G}}\sim \tau _{\text{H}}$ (cf. Sec. \ref%
{edistr}) and exponential distance growth rates, cf. Eq. (\ref{lgeqlh}), in
the gauge and Higgs sectors.

Nonetheless, the plaquette energy dependence of the MLEs in Figs. \ref%
{MLE1vsEp} and \ref{MLE2vsEp} also shows small systematic deviations from
linearity which become most notable towards the lowest $\bar{E}_{\text{p}}$
values. The same effect was observed in pure Yang-Mills theory \cite{mue96},
and a glance at the criterion (\ref{fslim}) indicates that finite-size
errors are responsible for the systematic upward trend of the MLEs at the
smallest $\bar{E}_{\text{p}}$. In fact, this is what one would intuitively
expect since field modes with longer average wavelengths are more strongly
deformed by the periodic boundary conditions. The slopes of the logarithmic
distance histories become most strongly modulated towards smaller $\bar{E}_{%
\text{p}}$ (cf. Sec. \ref{div}), furthermore, which introduces additional
systematic errors. Together with the finite-time errors to be discussed in
Sec. \ref{ltevol} they might cause additional deviations from a linear
energy dependence of the MLEs. Towards the maximal value $\bar{E}_{\text{p}%
}=4$ of the average energy per plaquette, on the other hand, lattice spacing
[cf. Eq. (\ref{lslim})] and compact phase-space artefacts are likely to
affect the results \cite{nie96,kra96}.

Since the YMH system has a second characteristic energy scale besides $\bar{E%
}_{\text{p}}$, i.e. the total energy $\bar{E}$ which additionally includes
both the energy stored in the Higgs field and in the gauge-Higgs
interactions (cf. Sec. \ref{edistr}) and is strictly conserved at all times,
it is natural to ask how the MLEs depend on $\bar{E}$. In order to answer
this question, we plot our MLEs in Figs. \ref{MLE1vsE} and \ref{MLE2vsE} as
a function of $\bar{E}$ (for the same $\bar{\kappa}$ and $N$ values as in
Figs. \ref{MLE1vsEp} and \ref{MLE2vsEp}) and find the dependence on the
total YMH\emph{\ }energy to be approximately linear as well:%
\begin{equation}
\bar{\lambda}_{\text{G},\text{H}}\left( \bar{E}\right) \simeq 0.55\bar{E}.
\label{lamEbar}
\end{equation}%
The above scaling behavior can be understood by recalling that our MLEs were
extracted during evolution times $t/a\leq 400$ over which the distances
generally saturate, but before the gauge and Higgs fields have started to
exchange appreciable amounts of energy. A glance at Fig. \ref{ens} shows
that after the gauge and Higgs sector have separately preequilibrated (i.e.
for $t/a\gg \bar{\lambda}^{-1}$), $\bar{E}_{\text{G}}=\bar{E}_{\text{p}}/2$
and $\bar{E}_{\text{H}}$ are practically time-independent in this phase.
Moreover, as mentioned in Sec. \ref{div}, staying in the weak-coupling
regime requires initial conditions which (except for the smallest $\delta
=0.05\,$) store considerably more energy in the gauge than in the Higgs
sector (cf. e.g. Fig. \ref{ens} which corresponds to $\delta =0.2$). In this
situation one derives from the definition of $\bar{E}$ in Sec. \ref{edistr},
which implies $\bar{E}_{\text{p}}=-4\bar{E}_{\text{H}}/3+10\bar{E}/3$, and
from Eq. (\ref{lamEp}) that 
\begin{equation}
\bar{\lambda}_{\text{G,H}}\left( \bar{E},\bar{E}_{\text{H}}\right) \simeq -%
\frac{2}{9}\bar{E}_{\text{H}}+\frac{5}{9}\bar{E}\overset{\bar{E}_{\text{H}%
}\ll \bar{E}}{\longrightarrow }\frac{5}{9}\bar{E},  \label{lamanalyt}
\end{equation}%
which explains the linear behavior and numerical slope of Eq. (\ref{lamEbar}%
). [Eq. (\ref{lamanalyt}) also explains the numerical scaling relation $\bar{%
\lambda}_{\text{G}}\left( \bar{E}\right) \simeq 0.53\bar{E}$ \cite{ful01}
for SU$\left( 2\right) $ YM theory where $\bar{E}_{\text{H}}\equiv 0$.] We
reemphasize that these results hold for MLEs extracted in the time window $%
\bar{\lambda}^{-1}\ll t/a\leq 400$ during which $\bar{E}_{\text{G}}$ and $%
\bar{E}_{\text{H}}\ll \bar{E}_{\text{G}}$ remain practically constant.

\subsection{ Long-time evolution of the Lyapunov histories}

\label{ltevol}

\begin{figure}[tbp]
\begin{center}
\includegraphics[height = 9cm]{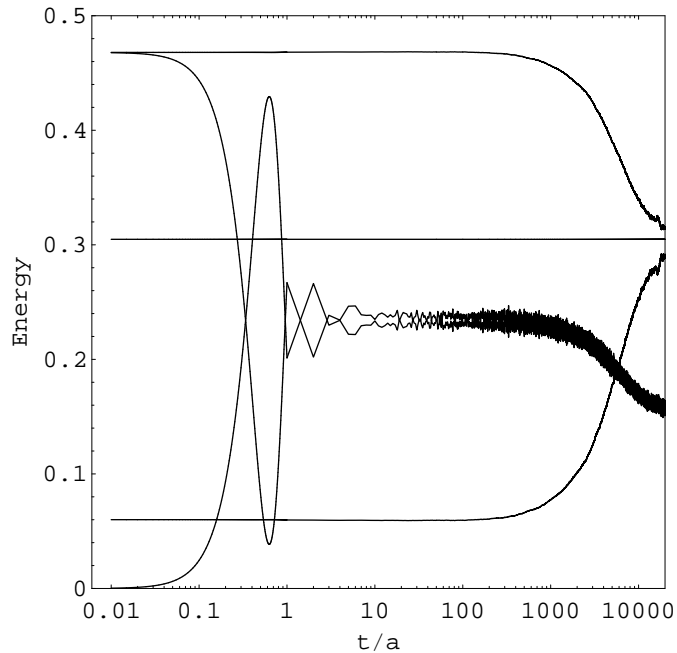}
\end{center}
\caption{Time evolution of the energy (per degree of freedom) stored in the
gauge field $E_{\text{G}}$ (uppermost initially horizontal line) and in the
Higgs field $E_{\text{H}}$ (lowermost initially horizontal line) and their
sum, i.e. the (conserved) total energy $E$ (horizontal line). The initially
oscillating line starting at zero is the electric gauge-field energy $E_{%
\text{el}}$, the one starting at the total initial gauge-field energy is the
magnetic (potential) gauge field energy $E_{\text{mag}}$. (The initial
conditions for the underlying trajectory were $\protect\delta =0.2$, $\bar{R}%
=0.2$, $\bar{\protect\kappa}=$ $1$, $N=10$ and $\Delta t=10^{-4}$.)}
\label{ens}
\end{figure}

\begin{figure}[tbp]
\begin{center}
\includegraphics[height = 9cm]{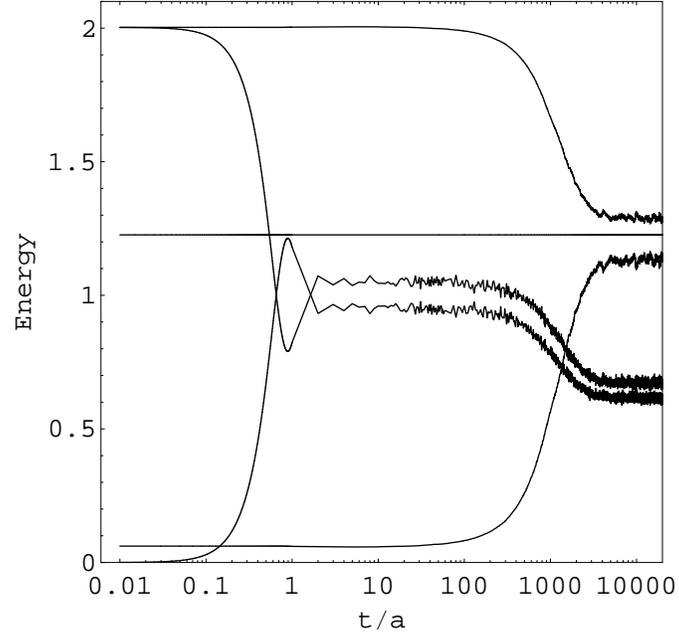}
\end{center}
\caption{Same as in Fig. \protect\ref{ens}, but for the trajectory subject
to the initial conditions $\protect\delta =1$, $\bar{R}=0.2$, $\bar{\protect%
\kappa}=$ $1$ and $N=10$.}
\label{ens2}
\end{figure}

\begin{figure}[tbp]
\begin{center}
\includegraphics[height = 21cm]{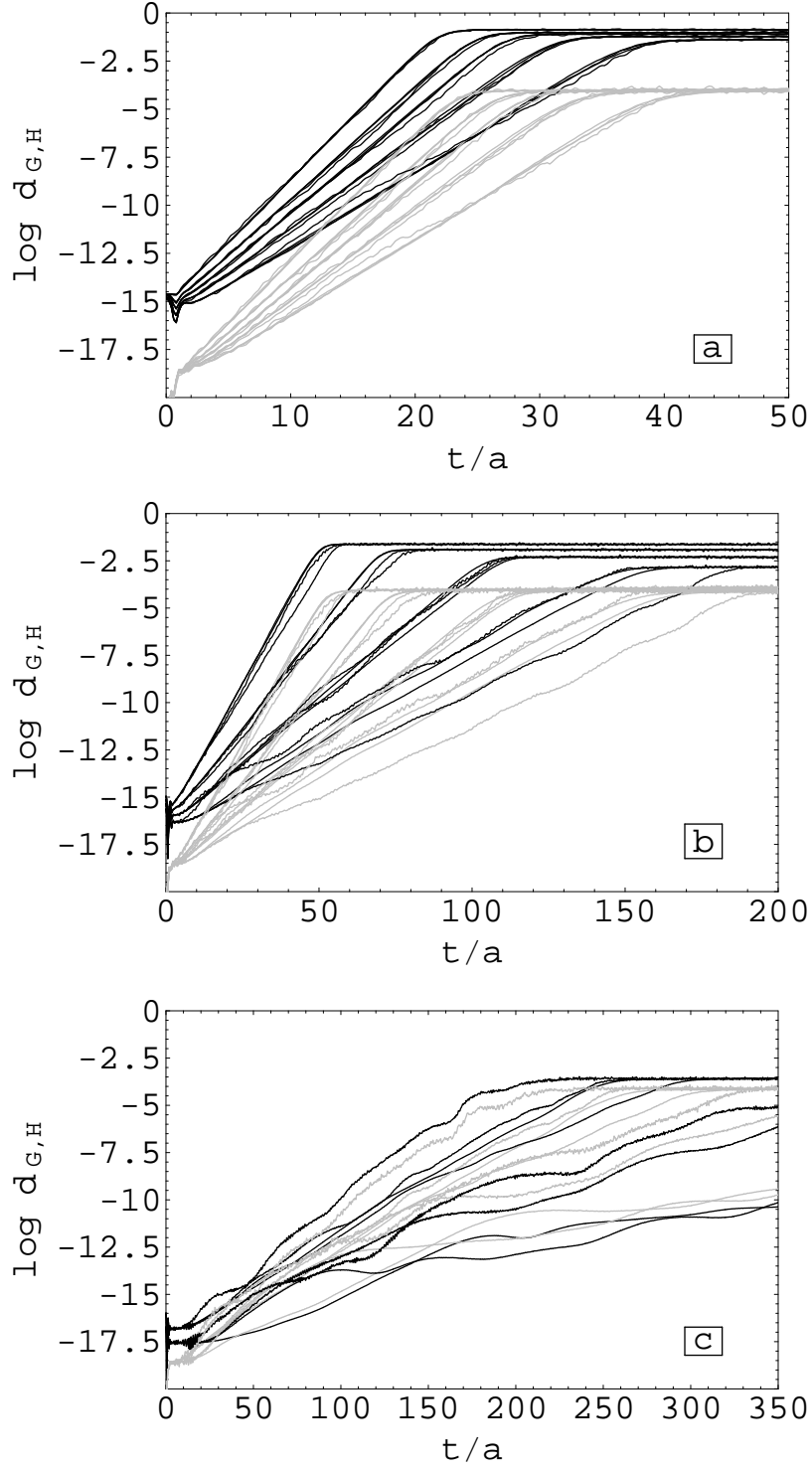}
\end{center}
\caption{The logarithmic distance evolution in the gauge (black) and Higgs
(grey) sectors at fixed Higgs self-coupling $\bar{\protect\kappa}=1$ is
plotted for four lattice volumina corresponding to $N=6,10,20$ and $30$. The
initial magnetic energy is parametrized by $\protect\delta $. Panel (a)
corresponds to $\protect\delta =1.0,0.5,0.45,0.4,0.35$, panel (b) to $%
\protect\delta =0.3,0.25,0.2,0.15$ and panel (c) to the two smallest values $%
\protect\delta =0.1,0.05$.}
\label{fig1}
\end{figure}

\begin{figure}[tbp]
\begin{center}
\includegraphics[height = 21cm]{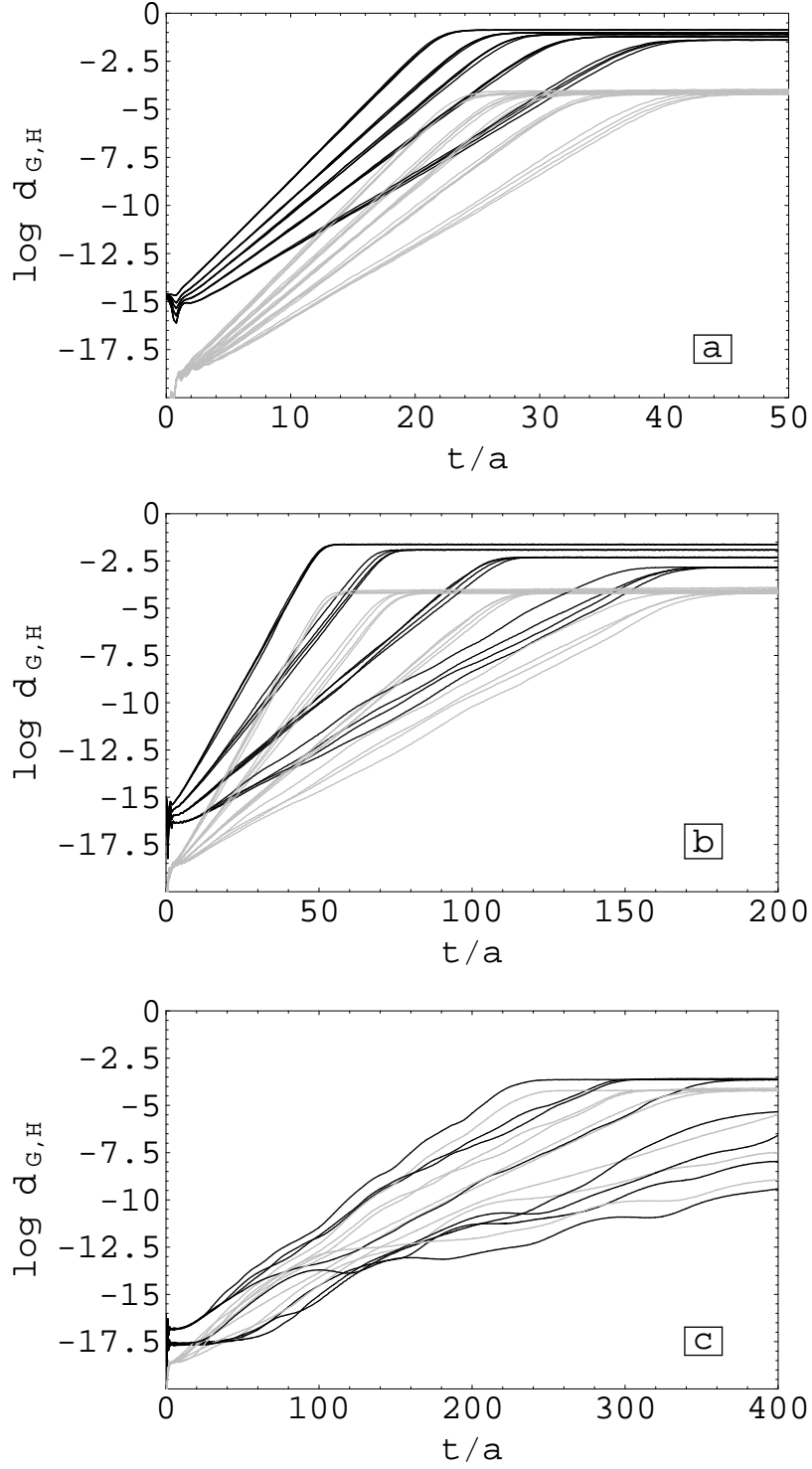}
\end{center}
\caption{The logarithmic distance evolution in the gauge (black) and Higgs
(grey) sectors on a $N=20$ lattice for four different Higgs self-couplings $%
\bar{\protect\kappa}=1,$ $8,$ $16$ and $24$. The distance trajectories are
grouped as in Fig. \protect\ref{fig1} according to their initial average
magnetic energy (parametrized by $\protect\delta $): panel (a) contains the
curves corresponding to $\protect\delta =1.0,0.5,0.45,0.4,0.35$, panel (b)
corresponds to $\protect\delta =0.3,0.25,0.2,0.15$ and panel (c) to the two
smallest values $\protect\delta =0.1,0.05$.}
\label{fig2}
\end{figure}

\begin{figure}[tbp]
\begin{center}
\includegraphics[height = 9cm]{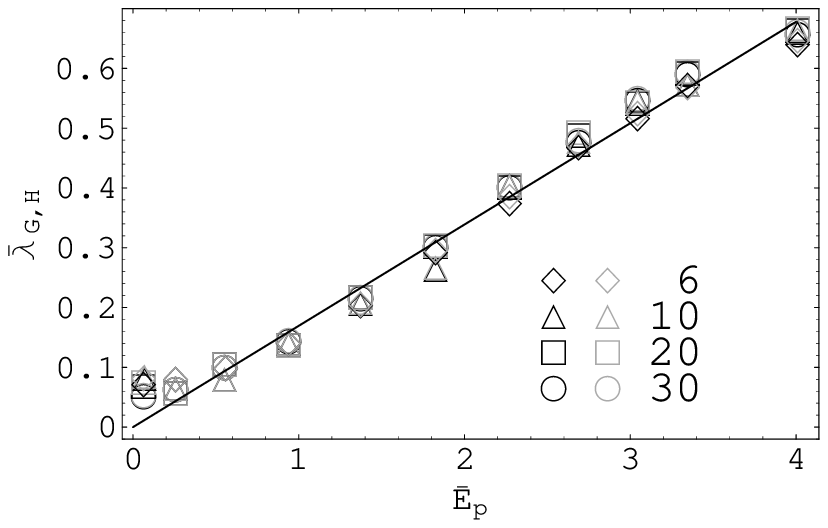}
\end{center}
\caption{Values of the maximal Lyapunov exponents $\bar{\protect\lambda}_{%
\text{G}}$ (black symbols) and $\bar{\protect\lambda}_{\text{H}}$ (grey
symbols) as a function of the average energy per plaquette $\bar{E}_{\text{p}%
}$ for $\bar{\protect\kappa}=1$ and $N=6,10,20,30$.}
\label{MLE1vsEp}
\end{figure}

\begin{figure}[tbp]
\begin{center}
\includegraphics[height = 9cm]{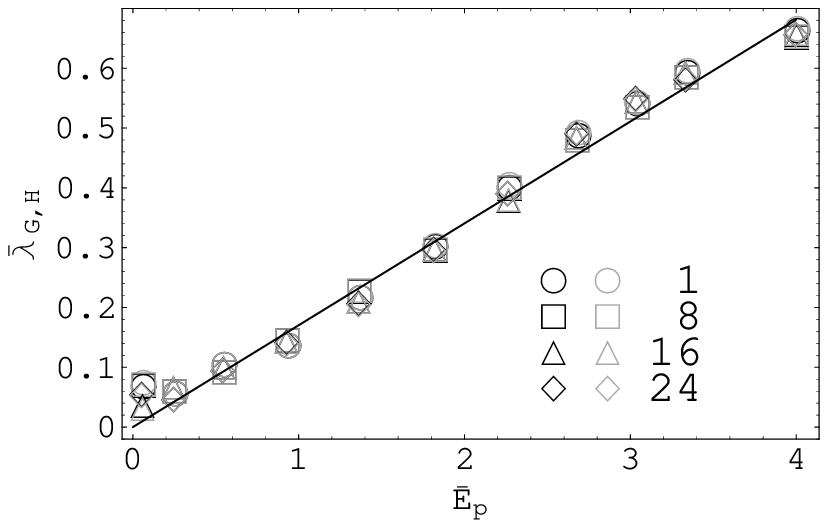}
\end{center}
\caption{Values of the maximal Lyapunov exponents $\bar{\protect\lambda}_{%
\text{G}}$ (black symbols) and $\bar{\protect\lambda}_{\text{H}}$ (grey
symbols) as a function of the average energy per plaquette $\bar{E}_{\text{p}%
}$ for $N=20$ and $\bar{\protect\kappa}=1,8,16,24$.}
\label{MLE2vsEp}
\end{figure}

\begin{figure}[tbp]
\begin{center}
\includegraphics[height = 9cm]{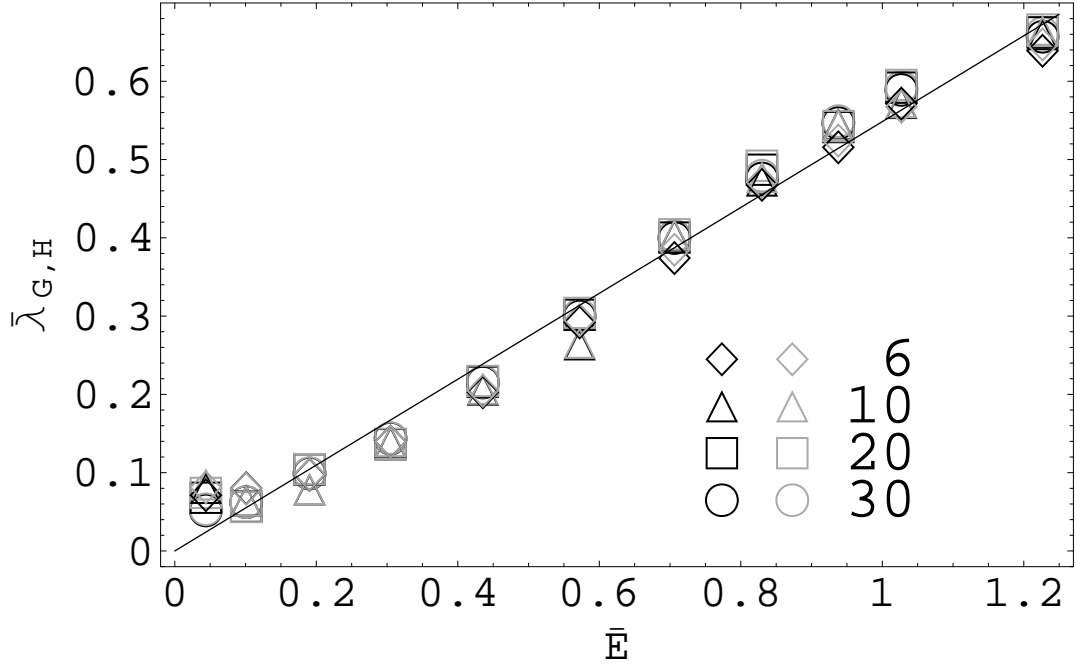}
\end{center}
\caption{Values of the maximal Lyapunov exponents $\bar{\protect\lambda}_{%
\text{G}}$ (black symbols) and $\bar{\protect\lambda}_{\text{H}}$ (grey
symbols) as a function of the total energy $\bar{E}$ for $\bar{\protect\kappa%
}=1$ and $N=6,10,20,30$.}
\label{MLE1vsE}
\end{figure}

\begin{figure}[tbp]
\begin{center}
\includegraphics[height = 9cm]{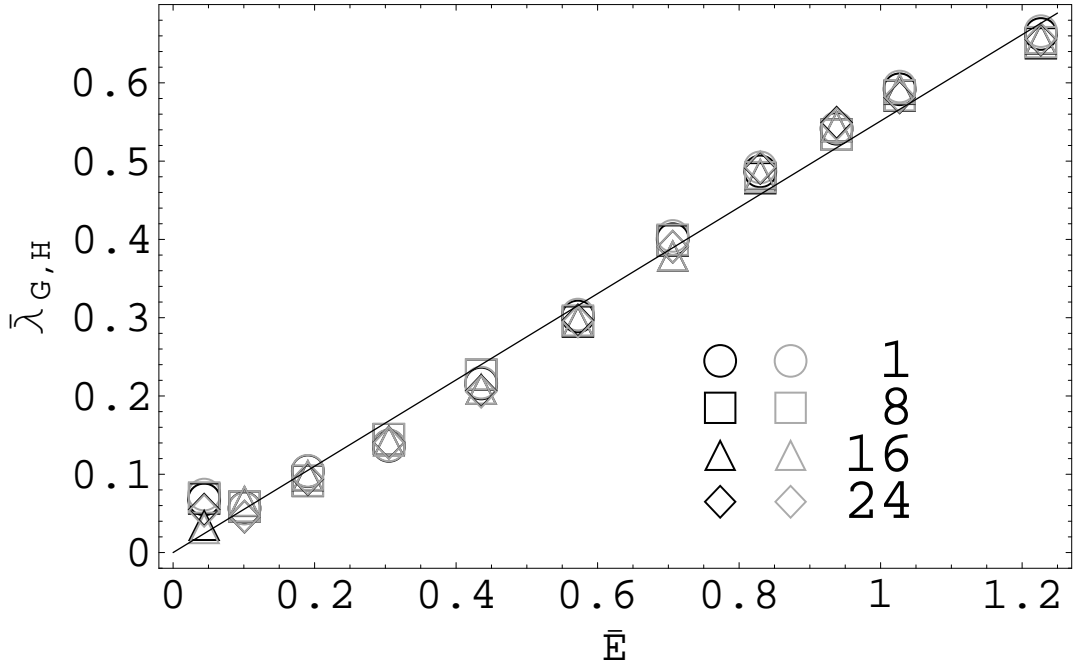}
\end{center}
\caption{Values of the maximal Lyapunov exponents $\bar{\protect\lambda}_{%
\text{G}}$ (black symbols) and $\bar{\protect\lambda}_{\text{H}}$ (grey
symbols) as a function of the total energy $\bar{E}$ for $N=20$ and $\bar{%
\protect\kappa}=1,8,16,24$.}
\label{MLE2vsE}
\end{figure}

At later times the Higgs sector will pick up energy from the gauge sector,
i.e. $\bar{E}_{\text{p}}$ will drop (for $\delta >0.05\,$) while $\bar{E}$
remains constant (cf. Figs. \ref{ens} and \ref{ens2}). In the $t\rightarrow
\infty $ limit the MLEs must attain a constant value, as implied in their
formal definition, and so will $\bar{E}_{\text{p}}$. This saturation is
strongly delayed, however, by the exceptionally long relaxation times which
govern the equilibration between the gauge and Higgs fields. In the
remainder of this section we will analyze the quantitative impact of this
saturation behavior on the extracted MLE values. To this end, we compute the
\textquotedblleft Lyapunov histories\textquotedblright\ \ 
\begin{equation}
\bar{\lambda}_{\text{G,H}}\left( \bar{E},\bar{\kappa},N;t\right) :=\frac{a}{t%
}\ln \frac{\tilde{d}_{\text{G,H}}^{\left( \text{alt}\right) }\left( \bar{E},%
\bar{\kappa},N;t\right) }{\tilde{d}_{\text{G,H}}^{\left( \text{alt}\right)
}\left( \bar{E},\bar{\kappa},N;0\right) }=\frac{a}{t}\sum_{k=1}^{t/\tau }\ln
s_{k}\overset{t\rightarrow \infty }{\longrightarrow }\bar{\lambda}_{\text{G,H%
}}\left( \bar{E},\bar{\kappa},N\right) 
\end{equation}%
[where $\tilde{d}_{\text{G,H}}^{\left( \text{alt}\right) }$ are the rescaled
distances (\ref{du2}) and (\ref{dh2}), and the $s_{k}$ are the rescaling
factors obtained after the $k$-th scaling step with rescaling period $\tau $%
], which approach the exact MLEs in the $t\rightarrow \infty $ limit, for
eight long-time field-pair trajectories in the time interval $t/a\in \left[
0,20000\right] $ on an $N=10$ lattice. (In order to improve numerical
efficiency, we increase the rescaling period $\tau $ with increasing
saturation time, i.e. with decreasing $\delta $, as detailed in the figure
captions below.)

\begin{figure}[tbp]
\begin{center}
\includegraphics[height = 9cm]{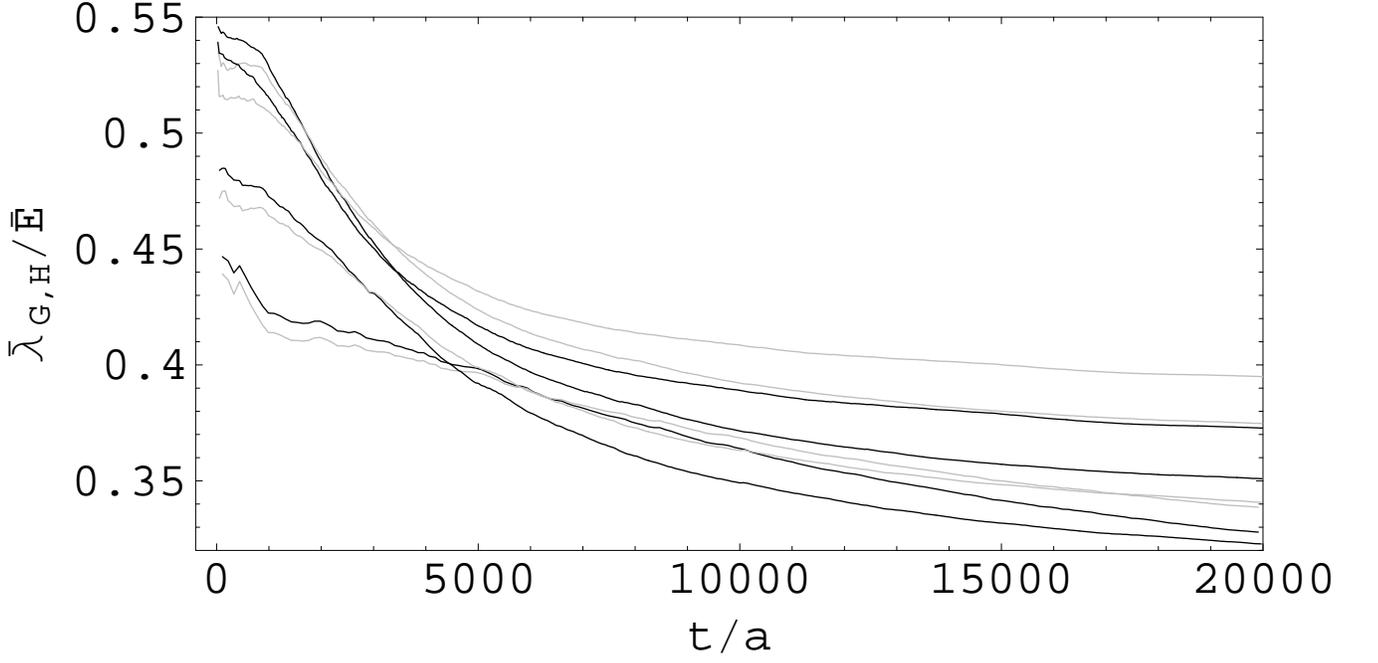}
\end{center}
\caption{Long-time evolution of the Lyapunov histories $\bar{\protect\lambda}%
_{\text{G}}\left( t\right) /\bar{E}$ (black lines)\ and $\bar{\protect\lambda%
}_{\text{H}}\left( t\right) /\bar{E}$ (grey lines) for $\left( \protect%
\delta ,\protect\tau \right) =\left( 0.6,25\right) $ (2nd largest initial
values), $\left( 0.45,30\right) $ (largest initial values), $\left(
0.3,55\right) $ (2nd lowest initial values) and $\left( 0.2,110\right) $
(lowest initial values) (with $\Delta t=0.0005,$ $\bar{\protect\kappa}=1,$ $%
\bar{R}=0.2$ and $N=10$).}
\label{lyetall}
\end{figure}

In Fig. \ref{lyetall} we show the Lyapunov histories $\bar{\lambda}_{\text{G}%
}\left( t\right) /\bar{E}$ (black lines)\ and $\bar{\lambda}_{\text{H}%
}\left( t\right) /\bar{E}$ (grey lines), \textquotedblleft
normalized\textquotedblright\ by the total energy, for the four long-time
field-pair histories with $\bar{\kappa}=1$ and initial magnetic energies
specified by $\delta =0.2,0.3,0.45$ and $0.6$. (This corresponds to
approximately equally spaced $\bar{E}_{\text{p}}$ values, cf. Figure \ref%
{epdel}.) A first important characteristic of all Lyapunov histories is
their monotonic decrease with time. Moreover, the saturation of $\bar{\lambda%
}_{\text{G}}\left( t\right) $ for large $t$ can be seen to proceed very
slowly: especially for smaller $\delta $ it is not fully completed even at $%
t=2\times 10^{4}a$. In all four cases, furthermore, $\bar{\lambda}_{\text{G}%
}\left( t\right) $ starts out somewhat larger than $\bar{\lambda}_{\text{H}%
}\left( t\right) $ but becomes smaller when the gauge and Higgs sectors
start to exchange substantial amounts of energy. The deviations between $%
\bar{\lambda}_{\text{G}}\left( t\right) $ and $\bar{\lambda}_{\text{H}%
}\left( t\right) $ remain at the one-percent level during the initial time
evolution (as reflected in the $\bar{\lambda}_{\text{G}}$ and $\bar{\lambda}%
_{\text{H}}$ estimates of Table 1) and increase systematically up to 5\%\ at 
$t=20000a$. Hence $\bar{\lambda}_{\text{G}}\left( t\right) $\ and $\bar{%
\lambda}_{\text{H}}\left( t\right) $ remain approximately equal over the
whole time evolution until they turn into the MLEs for $t\rightarrow \infty $%
.

\begin{figure}[tbp]
\begin{center}
\includegraphics[height = 9cm]{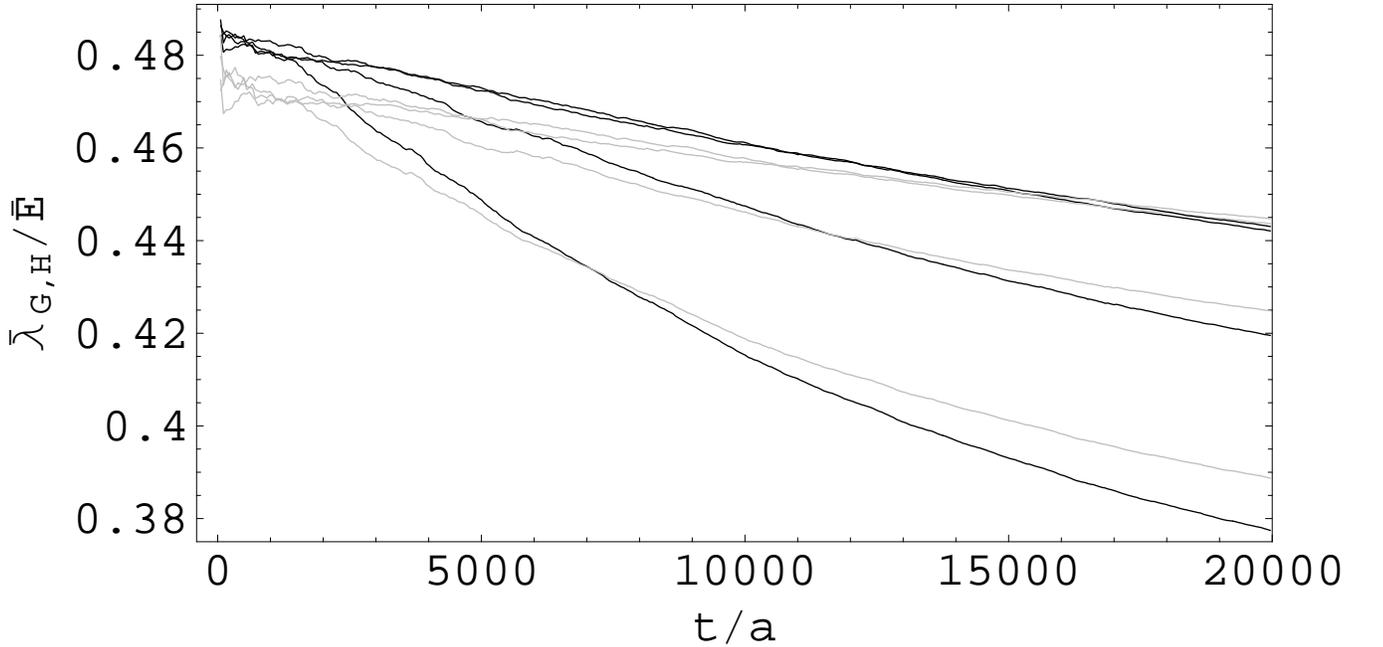}
\end{center}
\caption{Long-time evolution of the Lyapunov histories $\bar{\protect\lambda}%
_{\text{G}}\left( t\right) /\bar{E}$ (black lines)\ and $\bar{\protect\lambda%
}_{\text{H}}\left( t\right) /\bar{E}$ (grey lines) for $\bar{\protect\kappa}%
=8,$ $16,$ $24$ (from bottom to top) with $\Delta t=0.001$. The uppermost
curve, overlapping with its larger time-step counterpart, is for $\bar{%
\protect\kappa}=24$ with $\Delta t=0.0005$. (All curves correspond to $%
\protect\delta =0.3,\protect\tau =55,$ $\bar{R}=0.2$ and $N=10$).}
\label{lyetkap}
\end{figure}

In Fig. \ref{lyetkap} we display the Lyapunov histories $\bar{\lambda}_{%
\text{G,H}}\left( t\right) /\bar{E}$ for an intermediate $\delta =0.3$ and
the four values $\bar{\kappa}=8,16,24$ of the Higgs self-coupling. The main
tendencies observed in Fig. \ref{lyetall} remain intact for larger $\bar{%
\kappa}$, although increasing Higgs couplings further delay the saturation
of the Lyapunov histories. Indeed, already for $\bar{\kappa}=8$ it is more
difficult to reliably extrapolate $\bar{\lambda}_{\text{G,H}}\left( t\right) 
$ from the simulation interval $t/a\in \left[ 0,20000\right] $ to the MLE
value at $t\rightarrow \infty $. On the other hand, larger $\bar{\kappa}$
values further reduce the deviations between $\bar{\lambda}_{\text{G}}\left(
t\right) $ and $\bar{\lambda}_{\text{H}}\left( t\right) $ (which suggests
that the opposite tendency observed in Ref. \cite{bir96} was due to a
numerical artefact), and they also reduce the long-time variations of $\bar{%
\lambda}_{\text{G,H}}\left( t\right) /\bar{E}$ and hence the finite-time
errors of the MLEs.

In order to check whether the used integration time step $\Delta t=0.001$ is
small enough, we have also performed a simulation with half of its value for 
$\bar{\kappa}=24$. The corresponding curve, also drawn in Fig. \ref{lyetkap}%
, is essentially identical to the one with the larger time step, which shows
that the latter has no relevant time discretization error. Finally, we note
that when the Lyapunov histories decrease during equilibration,\ one may
expect their sensitivity to the Higgs sector to become larger. Figure \ref%
{lyetkap} shows that their dependence on the Higgs self-coupling $\bar{\kappa%
}$ is negligible at smaller evolution times ($t\lesssim 2000a$), as manifest
in Table 1, but indeed becomes more pronounced for larger $t$.

\begin{figure}[tbp]
\begin{center}
\includegraphics[height = 9cm]{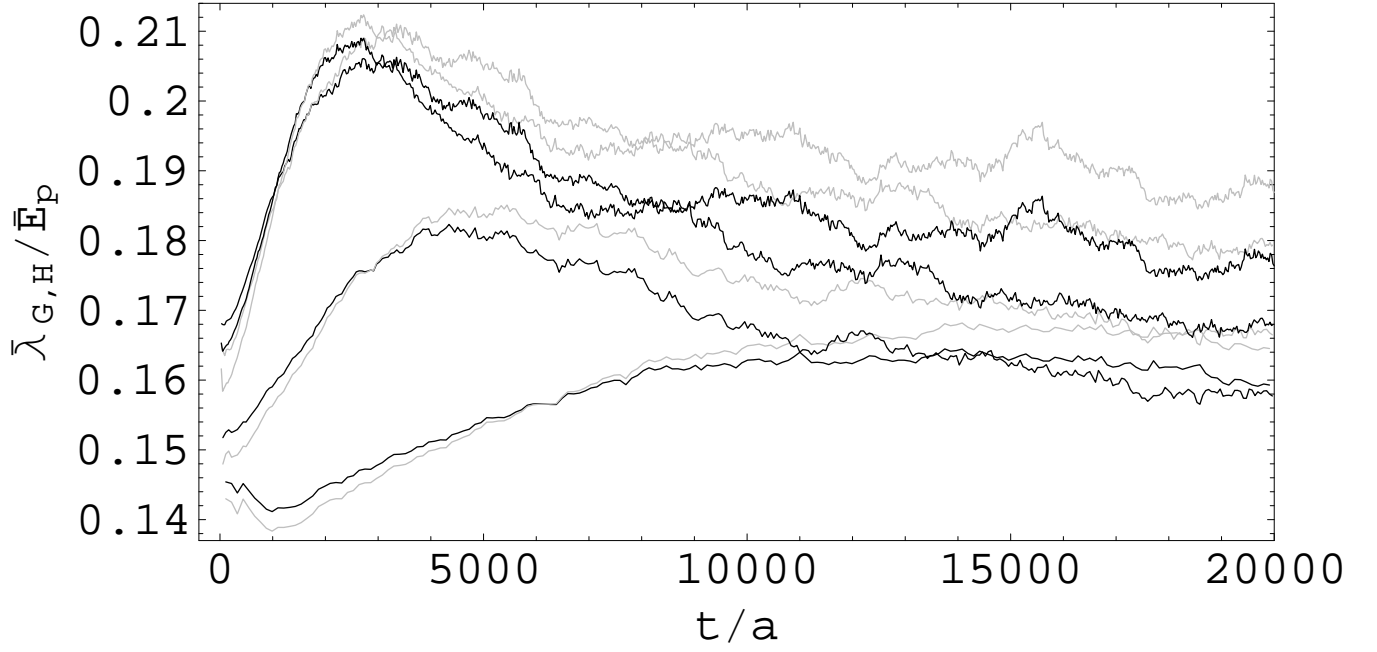}
\end{center}
\caption{Long-time evolution of the Lyapunov histories $\bar{\protect\lambda}%
_{\text{G}}\left( t\right) /\bar{E}_{\text{p}}$ (black lines)\ and $\bar{%
\protect\lambda}_{\text{H}}\left( t\right) /\bar{E}_{\text{p}}$ (grey lines)
for $\protect\delta =0.6$ (largest starting values), 0.45, 0.3 and $0.2$
(lowest starting values) (with $\bar{\protect\kappa}=1$ and $N=10$).}
\label{lyepall}
\end{figure}

\begin{figure}[tbp]
\begin{center}
\includegraphics[height = 9cm]{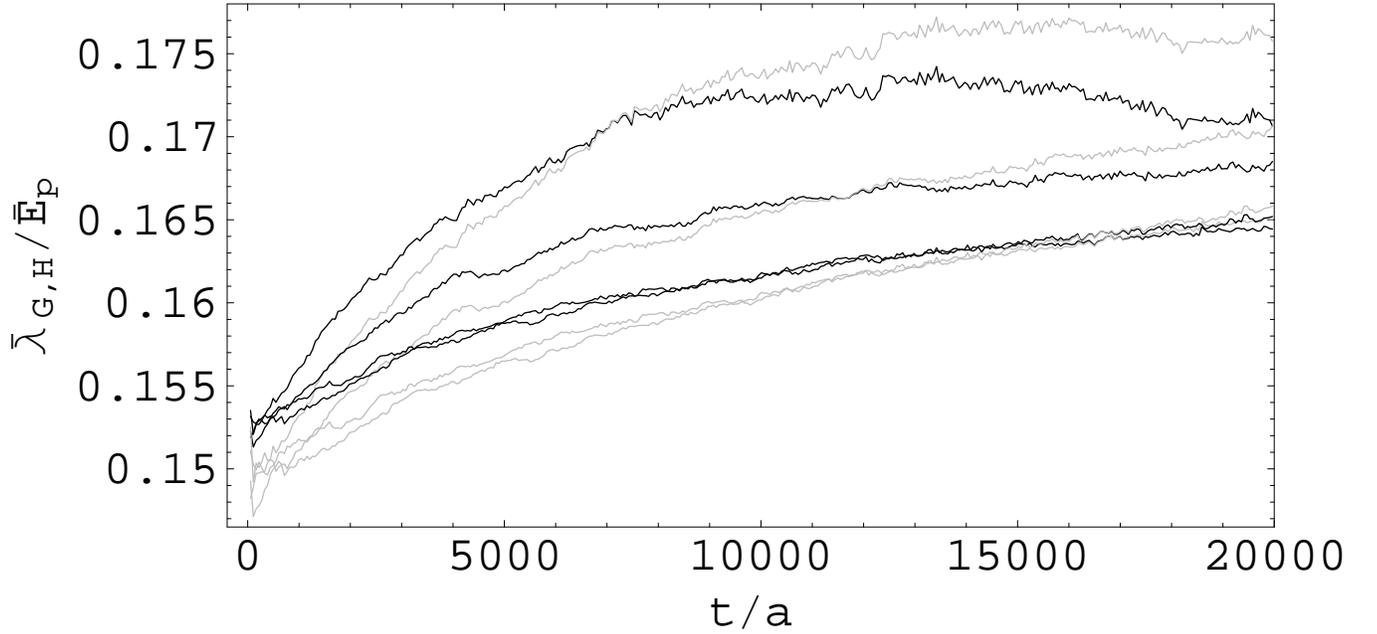}
\end{center}
\caption{Long-time evolution of the Lyapunov histories $\bar{\protect\lambda}%
_{\text{G}}\left( t\right) /\bar{E}_{\text{p}}$ (black lines)\ and $\bar{%
\protect\lambda}_{\text{H}}\left( t\right) /\bar{E}_{\text{p}}$ (grey lines)
for $\bar{\protect\kappa}=8,$ $16,$ $24$ and $\Delta t=0.001$ (from top to
bottom) and $\bar{\protect\kappa}=24$ with $\Delta t=0.0005$ ($\protect%
\delta =0.3,\protect\tau =55,$ $\bar{R}=0.2$ and $N=10$).}
\label{lyepkap}
\end{figure}

We now turn to the examination of the ratios $\bar{\lambda}_{\text{G}}\left(
t\right) /\bar{E}_{\text{p}}\left( t\right) $ and $\bar{\lambda}_{\text{H}%
}\left( t\right) /\bar{E}_{\text{p}}\left( t\right) $ which we plot in Figs. %
\ref{lyepall} and \ref{lyepkap} for the same parameter values as in Figs. %
\ref{lyetall} and \ref{lyetkap}. The closely parallel movement of $\bar{%
\lambda}_{\text{G}}\left( t\right) $ and $\bar{\lambda}_{\text{H}}\left(
t\right) $, and the systematics of their small deviations remain visible
here as well. The initial drop in $\bar{\lambda}_{\text{G,H}}\left( t\right)
/\bar{E}_{\text{p}}\left( t\right) $ (in particular for $\delta =0.2$) falls
into the time period during which $\bar{E}_{\text{p}}\left( t\right) $ is
practically constant, i.e. it is caused by the decrease of $\bar{\lambda}_{%
\text{G}}\left( t\right) $. Later on the gauge-field energy $\bar{E}_{\text{p%
}}\left( t\right) $ starts to drop (cf. Fig. \ref{ens}) and overcompensates
the continuing decrease of $\bar{\lambda}_{\text{G,H}}\left( t\right) $.
This causes the ratios $\bar{\lambda}_{\text{G,H}}\left( t\right) /\bar{E}_{%
\text{p}}\left( t\right) $ to rise. For $\delta =0.2$, one furthermore finds
from (slightly extrapolating) Fig. \ref{ens} that $\bar{E}_{\text{p}}\left(
t\right) =2\bar{E}_{\text{G}}\left( t\right) \rightarrow 2\bar{E}$ for $%
t\gtrsim 13000a$, so that the continuing, slight decrease of 
\begin{equation}
\frac{\bar{\lambda}_{\text{G,H}}\left( t\right) }{\bar{E}_{\text{p}}\left(
t\right) }\overset{t/a\gtrsim 13000,\text{ }\delta =0.2}{\longrightarrow }%
\frac{1}{2\bar{E}}\bar{\lambda}_{\text{G,H}}\left( t\right) 
\end{equation}%
for $t\gtrsim 13000a$ has again to be attributed solely to the behavior of $%
\bar{\lambda}_{\text{G,H}}\left( t\right) $. An important result, visible in
both Figs. \ref{lyepall} and \ref{lyepkap}, is that the ratios $\bar{\lambda}%
_{\text{G,H}}\left( t\right) /\bar{E}_{\text{p}}\left( t\right) $ saturate
significantly earlier than $\bar{\lambda}_{\text{G,H}}\left( t\right) /\bar{E%
}$ even at larger values of $\bar{\kappa}$. [Short-time fluctuations of the
average plaquette energy $\bar{E}_{\text{p}}\left( t\right) $ cause the time
evolution of $\bar{\lambda}_{\text{G,H}}\left( t\right) /\bar{E}_{\text{p}%
}\left( t\right) $ to appear more ragged.] This indicates that for large $t$%
, i.e. on the approach to full equilibrium, the average gauge energy $\bar{E}%
_{\text{p}}\left( t\right) /2$ decreases at the same rate as $\bar{\lambda}_{%
\text{G,H}}\left( t\right) $.

As already alluded to, our long-time evolution results allow for a
quantitative assessment of the finite-time errors in the MLE estimates of
Table 1, which were extracted at rather short evolution times. Figure \ref%
{lyetall} indicates that for $\bar{\kappa}=1$ the time variations of $\bar{%
\lambda}_{\text{G}}\left( t\right) $ reach about 25\% for $\delta =0.2$ and
about 30\% for $\delta =0.6$, while they remain about 5\% smaller for the
corresponding $\bar{\lambda}_{\text{H}}\left( t\right) $. These variations
may be considered as a (conservative)\ upper bound on the systematic
finite-time errors, in particular for larger $\delta $ and smaller $\bar{%
\kappa}$ values, and on the corresponding overestimates of the $\bar{\lambda}%
_{\text{G,H}}$ in Table 1. (The tendency\ to overestimate the MLEs when
extracting them at shorter evolution times was also noted in Refs. \cite%
{bir94,mue96}.)

\begin{figure}[tbp]
\begin{center}
\includegraphics[height = 9cm]{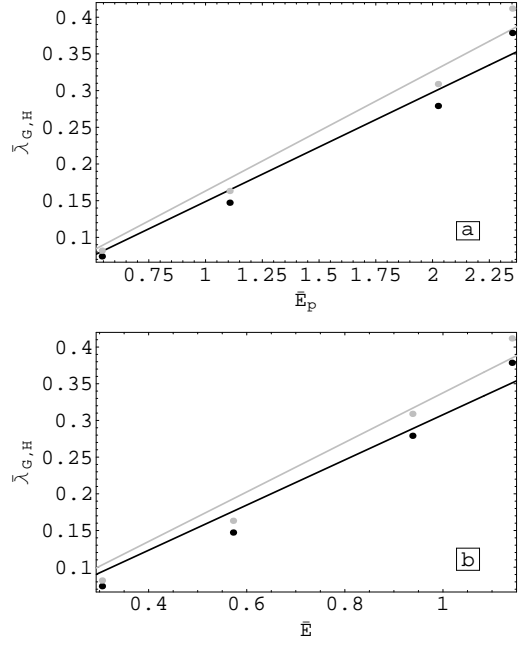}
\end{center}
\caption{Lyapunov histories $\bar{\protect\lambda}_{\text{G}}\left( t\right) 
$ (black dots)\ and $\bar{\protect\lambda}_{\text{H}}\left( t\right) $ (grey
dots), extrapolated to infinite evolution times, as a function of $\bar{E}_{%
\text{p}}$ (panel a) and $\bar{E}$ (panel b). The straight lines are the
best linear fits. (For $\protect\delta =0.6$, 0.45, 0.3 and $0.2$, with $%
\bar{\protect\kappa}=1$ and $N=10$.)}
\label{lmbdvsept}
\end{figure}

Finally, our long-time analysis allows us to clarify what happens to the two
scaling laws (\ref{lamEp}) and (\ref{lamEbar}) on the approach to total
equilibrium in the $t\rightarrow \infty $ limit where the Lyapunov histories
saturate. In fact, a rather reliable extrapolation to this limit (in
particular for larger energies)\ can be achieved by taking advantage of
empirical evidence for the asymptotic evolution-time dependence $\sim \bar{t}%
^{-1/2}$ with which the Lyapunov histories approach the MLEs \cite{mue96}.
(This finite evolution-time behavior is analogous to the finite-size
behavior $\sim \left( Na\right) ^{-1/2}$ \cite{mue96,ful01} which results
from sampling ergodic states \cite{bol00}. On similar grounds, the width of
Gaussian fluctuations around the mean value of the average Lyapunov exponent
was argued to decay as $\bar{t}^{-1/2}$ \cite{bol00}.) Hence the new
functions $\bar{\lambda}_{\text{G,H}}\left( \bar{t}^{-1/2}\right) $ can
approximately be fitted by straight lines (with a potential systematic bias)
and the MLEs determined as the intersections with the $\bar{t}^{-1/2}=0$
axis \cite{mue96}. Extrapolating the curves in Figs. \ref{lyetall} and \ref%
{lyetkap} in this way to infinite evolution time yields our best estimates
for the MLEs which we plot in Fig. \ref{lmbdvsept} as a function of $\bar{E}%
_{\text{p}}$ (upper panel) and $\bar{E}$ (lower panel). These figures show
that both dependencies remain to good accuracy linear. The best linear fits
(also shown in the figures) are%
\begin{equation}
\bar{\lambda}_{\text{G}}\left( \bar{E}_{\text{p}}\right) \simeq 0.149\bar{E}%
_{\text{p}},\text{ \ \ \ \ \ }\bar{\lambda}_{\text{H}}\left( \bar{E}_{\text{p%
}}\right) \simeq 0.163\bar{E}_{\text{p}},  \label{lamEpAsym}
\end{equation}%
and%
\begin{equation}
\bar{\lambda}_{\text{G}}\left( \bar{E}\right) \simeq 0.308\bar{E},\text{ \ \
\ \ \ }\bar{\lambda}_{\text{H}}\left( \bar{E}\right) \simeq 0.338\bar{E}.
\label{lamEbarAsym}
\end{equation}%
Equations (\ref{lamEpAsym}) show that to an accuracy of at least about 10\%
the scaling law (\ref{lamEp}) found before full equilibration, with the same
coefficient as in pure YM theory, indeed remains intact asymptotically.
(First indications for this behavior were observed in Ref. \cite{hei97} on
the basis of a trajectory for $t\leq 12000a$.) Qualitatively, this is also
reflected in Fig. \ref{lyepall} where the ratios $\bar{\lambda}_{\text{G,H}%
}\left( t\right) /\bar{E}_{\text{p}}\left( t\right) $ start at around $1/6$
and asymptotically return to it for very large $t$ (while in the meantime
deviating by maximally (i.e. for the largest $\delta $) about 20\%, mainly
when most of the energy is exchanged between the gauge and Higgs fields).

The Eqs. (\ref{lamEpAsym}) also explain the $\bar{E}$ dependence of the
asymptotic Lyapunov histories exhibited by the fits (\ref{lamEbarAsym}).
Indeed, after full equilibration at $t\rightarrow \infty $ with $\bar{E}_{%
\text{G}}=\bar{E}_{\text{p}}/2=\bar{E}_{\text{H}}=\bar{E}$ one expects from $%
\bar{\lambda}_{\text{G,H}}\left( \bar{E}_{\text{p}}\right) =\alpha _{\text{%
G,H}}\bar{E}_{\text{p}}\simeq \bar{E}_{\text{p}}/6$ that 
\begin{equation}
\bar{\lambda}_{\text{G,H}}\left( \bar{E}\right) =2\alpha _{\text{G,H}}\bar{E}%
\simeq \frac{1}{3}\bar{E}  \label{lamEbaranalyt}
\end{equation}%
which is within errors identical to Eqs. (\ref{lamEbarAsym}). Hence in
equilibrium the linear dependence of the MLEs on $\bar{E}_{\text{p}}$
implies a linear dependence on $\bar{E}$ with twice the slope (as realized
to good accuracy in the fits (\ref{lamEpAsym}) and (\ref{lamEbarAsym})).
This fact went probably unnoticed in Ref. \cite{hei97} which argued against
linear scaling of the $\bar{\lambda}_{\text{G,H}}$ with $\bar{E}$ on the
basis of field evolution trajectories over maximally several thousand
lattice time units, i.e. likely too short to bring the system close enough
to equilibrium. (In addition, our above observation supports the diagnosis
of Ref. \cite{ful01} which attributes the logarithmic energy dependence
found numerically for several SU$\left( 2\right) $-YM MLEs after very long
evolution times to finite-size artefacts of the monodromy-matrix method.)
More generally, linear scaling of $\bar{\lambda}_{\text{G,H}}$ with $\bar{E}%
_{\text{p}}$ implies a linear dependence on $\bar{E}$ in any time window
during which $\bar{E}_{\text{p}}\propto \bar{E}$. This condition seems to be
satisfied only when the gauge and Higgs sectors do not exchange relevant
amounts of energy, however, i.e. only in the preequilibration phase and
after mutual equilibration is essentially achieved. Nevertheless, the slopes
of the $\bar{E}$ dependence in these two time intervals are different (5/9
and 1/3, respectively).

The above evidence for the linear dependence (\ref{lamEp}) of the MLEs on
the average magnetic energy $\bar{E}_{\text{p}}$ to prevail for $%
t\rightarrow \infty $ appears consistent with our previous indications for
the maximally chaotic mode to reside mainly in the gauge sector, with our
finding that the scaling behavior (\ref{lamEp}) sets in way before the gauge
fields have full access to the energy stored in the Higgs sector, and with
the result that the ratios $\bar{\lambda}_{\text{G,H}}\left( t\right) /\bar{E%
}_{\text{p}}\left( t\right) $ saturate significantly faster than the $\bar{%
\lambda}_{\text{G,H}}\left( t\right) $ themselves.

\subsection{Maximal Lyapunov exponents of initially homogeneous magnetic
fields}

\label{mleb}

In this section we digress from our main subject and apply some of the
numerical techniques developed above to the time evolution of spacially
constant, non-Abelian magnetic fields. In the pioneering days of QCD such
homogeneous magnetic fields were perturbatively established to be unstable
in pure YM theory \cite{sav77}. This so-called Savvidy instability\ was
later explored in the nonperturbative domain by numerical methods similar to
ours \cite{tra92,mul92}. It provided early indications for the complexity of
the Yang-Mills vacuum and\ has triggered the development of stochastic and
chaotic concepts for vacuum structure and quark confinement \cite%
{bir294,far05,chaoconf}. In the following we are going to study the impact
of the matter (i.e. Higgs) fields on the Savvidy instability.

\begin{figure}[tbp]
\begin{center}
\includegraphics[height = 9cm]{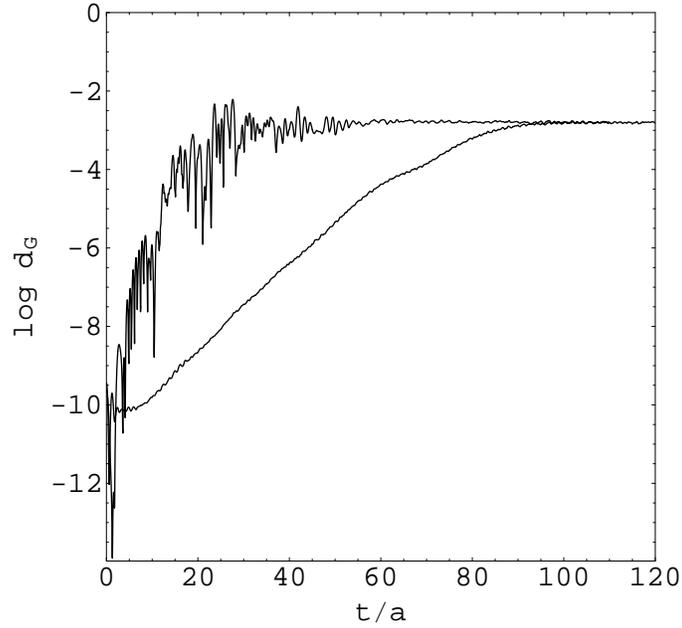}
\end{center}
\caption{Logarithmic distance evolution for initially neighboring,
homogeneous, non-Abelian magnetic fields (ragged curve) and for initially
neighboring randomized gauge-field configurations (smooth curve) in YM
theory (with $N=10$ and $\bar{E}=0.572023$).}
\label{cbYM}
\end{figure}

As a benchmark for comparison with the YMH case, we first reproduce the
nonperturbative time evolution of the distance (\ref{du}) between initially
adjacent, homogeneous magnetic fields in pure YM theory on an $N=10$
lattice. The non-Abelian magnetic field is defined as $B_{\text{p}}=\arccos
tr\left( U_{\text{p}}\right) $, and the fields are initialized with total
energy $\bar{E}=0.57$ by setting $E_{x,i}^{a}\left( 0\right) =0$, $%
B_{x,12}\left( 0\right) =0.899,$ $B_{x,13}\left( 0\right) =0.791$ and $%
B_{x,23}\left( 0\right) =1.453$ for all $x$. In Fig. \ref{cbYM} we compare
their logarithmic distance evolution to that of an initially randomized
gauge field (cf. Sec. \ref{init}) with the same energy. The constant
magnetic field pair turns out to have about twice the average slope of $\ln
d_{\text{G}}\left( t\right) $ in the linear region, i.e. the homogeneous
magnetic field is substantially less stable than the random field. This
result corroborates similar findings in Ref. \cite{mul92}.

\begin{figure}[tbp]
\begin{center}
\includegraphics[height = 9cm]{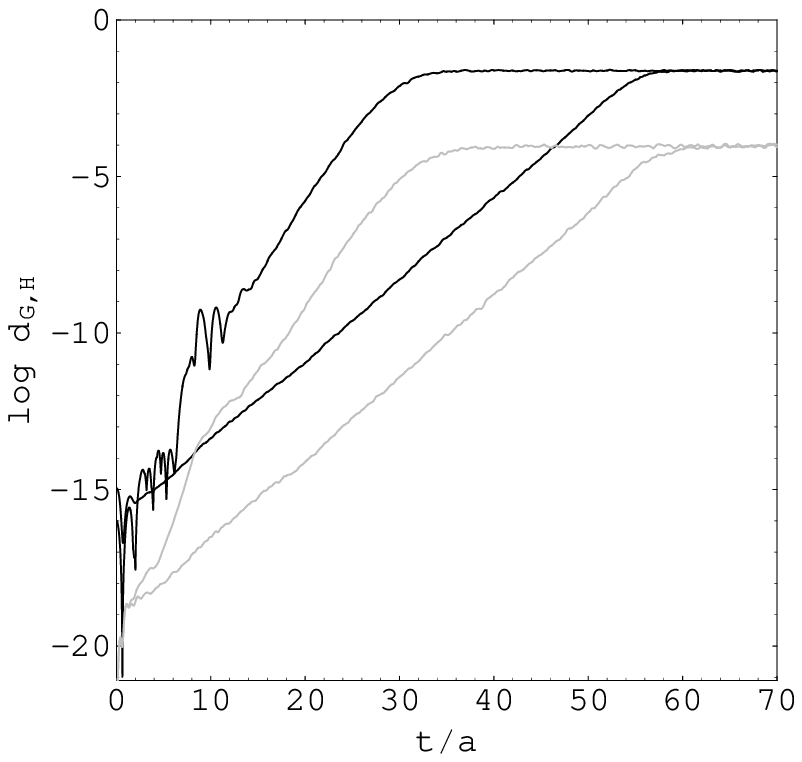}
\end{center}
\caption{Evolution of the logarithmic distances (\protect\ref{du}) (black
lines) and (\protect\ref{dh}) (grey lines) for two initially neighboring,
homogeneous, non-Abelian magnetic fields (ragged curves with larger slopes
in the linear region) and for two initially neighboring random gauge-field
configurations in YMH theory (with $\protect\delta =0.3$, $\bar{R}=0.2$, $%
\bar{\protect\kappa}=1$, $N=10$, $\bar{E}=0.572023$).}
\label{cbYMH}
\end{figure}

We now turn to the analogous time evolution of initially constant magnetic
fields in YMH theory, again on a lattice with $N=10$ sites per dimension. As
in all previous sections, the Higgs field is initialized at the spacially
constant value $\bar{R}_{x}\left( 0\right) =\bar{R}=0.2$, and the initial
values $\dot{\phi}\left( 0\right) =0$, $E_{x,i}^{a}\left( 0\right) =0$ are
imposed in order to\ satisfy Gauss' law (\ref{Gss}). The $B$ field is
initialized at the values $B_{x,12}\left( 0\right) =2.319,$ $B_{x,13}\left(
0\right) =2.152$ and $B_{x,23}\left( 0\right) =1.428$ for all $x$. We
further choose $\delta =0.3$ and $\bar{\kappa}=1$ in order to inject the
same total energy $\bar{E}=0.57$ as in the YM case above. The logarithmic
distance evolution\ under these conditions is displayed in Fig. \ref{cbYMH}
for the gauge and Higgs field metrics (\ref{du}) and (\ref{dh}), again
together with its counterpart for a corresponding random field. As in the
YM\ case, the homogeneous magnetic field produces about twice the slope in
the linear region. Hence the presence of the matter fields seems neither to
dampen nor to enhance the increased instability of the homogeneous magnetic
field configurations relative to a random configuration of the same total
energy. This seems to be consistent with our above evidence for the
chaoticity of YMH theory to be dominated by the gauge sector.

The slopes of $\ln d_{\text{G}}\left( t\right) $ for both constant magnetic
and random fields, however, are (in the linear region) about twice as large
in the YMH example of Fig. \ref{cbYMH} than in YM theory (Fig. \ref{cbYM}).
Probably this result depends rather strongly on the initial conditions, and
especially on how they distribute the initial energy over the gauge and
Higgs field sectors. Our above initial conditions were chosen to provide a
demonstrative example for the presence of matter fields to strongly enhance
the Savvidy instability. This massive impact raises the possibility that
simulation results for those gauge-field instabilities which drive
isotropization and thermalization in the aftermath of high-energy nuclear
collisions could be significantly modified in the presence of quark fields
as well (although they have smaller occupation numbers).

\section{Related equilibration processes in cosmology and nuclear collisions}

\label{impl}

In the following section we are going to discuss several aspects of
nonequilibrium processes in the early Universe and in the aftermath of
high-energy nuclear collisions which are pertinent in our context. We
comment on the impact of the chaotic thermalization properties calculated
above and on results of classical gauge-theory simulations related to ours.
We also suggest a few promising extensions of our work which would help to
clarify the role of chaotic thermalization processes after nuclear
collisions and the nature of the \textquotedblleft
apparent\textquotedblright\ or \textquotedblleft pre-\textquotedblright\
equilibrium at the beginning of the subsequent hydrodynamic evolution phase.

\subsection{Early Universe}

According to the inflation paradigm, the vacuum energy of one or more
classical, scalar inflaton fields dominated the very early Universe. This
energy caused a typically exponentially accelerated expansion period which
very efficiently diluted particles and fluctuations \cite{inf}. It left the
universe in a supercooled, highly nonthermal state which was practically
devoid of matter, radiation and entropy. Hundreds of models for the
phenomenologically very successful inflationary scenario were proposed \cite%
{bas06}. A compelling \textquotedblleft microscopic\textquotedblright\
theory, however, which is able to explain e.g. the nature of the inflaton(s)
and the very specific properties of their dynamics, has not yet been
established, partly because it involves unknown physics beyond the standard
model.

For the post-inflationary reheating period, during which the Universe
thermalized at a still very large \textquotedblleft reheating
temperature\textquotedblright , the theoretical situation is similar:\ there
exist many scenarios and model calculations for specific processes whereas
the underlying dynamics as a whole is not yet settled. During reheating a
huge amount of entropy was released. All the matter and radiation of the
present Universe was created and the energy density of the inflaton(s) was
transformed into a hot and ultrarelativistic plasma. (Afterwards the
Universe expanded essentially in the Friedman-Robertson-Walker geometry and
cooled almost isoentropically according to the \textquotedblleft
hot-big-bang\textquotedblright\ scenario.) In the following we will be
particularly interested in the (semi-) classical phases during the reheating
period, when the large occupation numbers of the participating field modes
made contributions from chaotic thermalization relevant.

The probably most important of these phases, referred to as preheating, is
suggested to have taken place immediately after inflation \cite{kof94}.
During this very short period, which lasted about $10^{-35}$ secs., particle
production became explosive. Preheating can be induced either by a tachyonic
instability of the inhomogeneous modes which accompany electroweak symmetry
breaking \cite{fel01}, or by the stimulated decay of an almost homogeneous
inflaton which coherently oscillates with an initial amplitude\ of the order
of the Planck mass. In the latter case, the accelerated decay is the
consequence of a parametric resonance with condensates composed of the
produced particles \cite{kof94}. A detailed understanding of the preheating
process is particularly crucial because the bulk of the initial conditions
for the subsequent thermal history of the Universe are settled at its end.
Since the homogeneous energy density of the inflaton transfers exponentially
rapidly into highly occupied, inhomogeneous out-of-equilibrium modes,
furthermore, the violently nonperturbative and\ nonequilibrium processes
underlying preheating are amenable to classical lattice simulations \cite%
{mic04}.

After preheating, a variety of crucial thermalization processes began to
drive the Universe towards equilibrium. Classical lattice simulations of
such processes indicate, furthermore, that the infrared modes excited during
preheating evolve towards a saturated occupation number distribution long
before thermalization completes \cite{pod06}. Such effects have been
interpreted as signs of \textquotedblleft
prethermalization\textquotedblright , characterized by an energy-pressure
relation approximating an equation of state \cite{ber04,pod06}. Later on,
the distributions move towards complete saturation by cascading towards
ultraviolet and infrared modes (as in Kolmogorov wave turbulence). During
the last and longest stage of equilibration, finally, the particle
distributions become fully thermal \footnote{%
Full thermalization has to happen fast enough and at low enough temperatures
to prevent the previously generated baryon asymmetry to be washed out by
sphaleron transitions.}. Simultaneously, the occupation numbers drop until
quantum physics eventually dominates and classical simulations become
ineffective.

Among the many important nonequilibrium processes which shape the reheating
period are backreactions on the inflaton field which eventually stop
particle production, violent and still nonperturbative particle rescattering
events which very efficiently generate entropy, the nonthermal production of
heavy particles as well as phase transitions. Further reheating processes
which left crucial imprints in our present Universe are primordial magnetic
field generation \cite{dia08}, topological and large-scale structure
formation as well as baryogenesis \cite{pro97}, which created the observed\
abundance of baryons over antibaryons. In our context, baryogenesis is
particularly interesting since it can efficiently proceed only far from
equilibrium. As already mentioned, lattice simulations of classical
real-time field evolution are a method of choice for the analysis of such
processes. In fact, they turned out to be particularly useful for shedding
light on the preheating phase whose immense particle production rate and
condensate formation require a fully nonperturbative treatment while the
generated, large (boson)\ occupation numbers ensure a classical field
evolution \cite{preh}.

In the following, we will focus on classical lattice simulations of
cosmological pre- and reheating processes which were based on dynamics
including the gauge-scalar sector of the standard model \footnote{%
The YMH model should provide a reasonable description of standard model
physics during the early evolution phases in which quark contributions are
small.}, i.e. the\ SU$\left( 2\right) $\ YMH theory with a fundamental Higgs
doublet (\ref{ham}) \footnote{%
We are grateful to Anders Tranberg for bringing some of these studies to our
attention.}. This theory has been used, notably, to explore the electroweak
symmetry breaking transition \cite{esb}. It underlies our own work,
furthermore, and thus allows for some partial and qualitative comparisons --
although the initial conditions (and the additional field content including
e.g. an inflaton) required to describe cosmological situations differ
considerably from the random ones which we have adopted above.

The analysis of electroweak baryogenesis at energies of the order of 100 GeV
is, as already alluded to, an especially interesting application of the YMH
model (\ref{ham}) and its extensions \cite{kuz85,pro97}. Such studies are
mainly motivated by the question whether \textquotedblleft minimal
extensions\textquotedblright\ of the electroweak standard model, which
implement e.g. additional neutral scalar inflaton field(s) or $CP$\
violating couplings, are able to explain the observed baryon asymmetry of
the present Universe. As a step towards clarifying this issue, the baryon
(and lepton) number and $CP$ violating sphaleron transition rate and the
Chern-Simons number diffusion in the unbroken phase were studied in Refs. 
\cite{amb91,tan96,amb97,moo99}. In a hybrid inflation scenario \cite{gar99},
based on an additional singlet inflaton which couples to the Higgs field to
accelerate particle production (compared to only gravitational coupling),
the nonequilibrium preheating dynamics was found to generate Chern-Simons
number, a prerequisite for electroweak baryogenesis, locally and
stochastically \cite{gar04}. (For some cautionary remarks on the reliability
of classical lattice results in this preheating scenario for baryogenesis
see Ref. \cite{moo01}.) In the same dynamical framework, primordial magnetic
fields are produced with sufficient magnitude and correlations to act as
seeds for the magnetic fields observed in galaxies and galaxy clusters today 
\cite{dia08}. Electroweak baryogenesis during a cold electroweak transition
with tachyonic preheating (induced by a spinodal Higgs field instability)
and additional $CP$ violation generated by a coupling of the Higgs field to
the topological charge density of the gauge field, has been investigated in
Ref. \cite{tra03}. The particle distribution functions of the Higgs and
gauge fields (on which kinetic theory is based) were extracted in Ref. \cite%
{sku03} from the correlators of the simulated classical fields, with the
electroweak phase transition modeled by a quench.

Our main goal in the present paper was to study generic chaotic
thermalization properties of YMH theory on the basis of random initial
conditions. In this respect, our work differs from the more specialized
simulations discussed above (which partly also include additional dynamics).
Although this prevents a quantitative comparison of the results, we believe
that the chaotic thermalization processes which we have analyzed should be
relevant for most of the mentioned post-inflationary pre- and reheating
processes as well. By adapting the initial conditions to cosmological
situations, in particular, one could directly investigate the contributions
of deterministic chaos, as measured by the Lyapunov exponents, to specific
nonequilibrium processes. It would for example be interesting to follow the
sphaleron transition and magnetic field production rates during the
different stages of chaotic thermalization. It would also be useful to study
the impact of the lattice spacing on the chaotic thermalization rates \cite%
{moo01} and to include physics beyond the standard model, e.g. an inflaton
field, into the analysis.

\subsection{Nuclear collisions}

Thermalization properties of excited quark-gluon matter, as produced at the
SPS, RHIC, LHC and (future) FAIR colliders, have been intensely studied in
various approaches of increasing sophistication \footnote{%
The earlier approaches, based on perturbative QCD, include several Monte
Carlo parton cascade models which later took gluon radiation in the initial
and final stages of the parton interactions into account \cite{cas},
phenomenological nonperturbative string interactions, multiparticle
processes with specific (physical) infrared cutoffs for the parton cross
section (which speed up equilibration but depend crucially on details of
their definition and regularization) etc. Numerical implementations of
transport theory approaches (Boltzmann and Fokker-Planck equations, with
multiparticle processes based e.g. on Schwinger's particle creation
mechanism) were also studied, and analytical progress was made e.g. by
showing that gluons equilibrate much faster than quarks \cite{shu92}.} (see
Refs. \cite{mro206} for recent reviews). The detailed local equilibration
mechanisms are of central importance for the heavy-ion programs since they
determine whether a new, deconfined state of matter -- the quark-gluon
plasma -- can be locally thermalized in the aftermath of high-energy nuclear
collisions, i.e. whether the produced system equilibrates fast enough for
thermodynamic concepts to apply before it disintegrates.

The thermalization issue became even more intriguing when RHIC results
showed that the produced matter\ starts to behave collectively after times
of less than $1$ fm/$c$\ and is subsequently described by essentially\ ideal
Bjorken hydrodynamics with almost maximal elliptic flow \cite{rhic}. These
findings are generally interpreted as a surprisingly\ fast apparent\
thermalization of the system, which minimally requires the isotropization of
the long-wavelength modes participating in the\ hydrodynamic behavior\ \cite%
{arn205} and perhaps the onset of prethermalization \cite{ber04}. In any
case, the very short (pre-) equilibration time cannot be explained by weakly
coupled parton-parton collisions alone \cite{mro206,arn03}.

The time evolution of a typical RHIC reaction (which is sometimes referred
to as a \textquotedblleft little bang\textquotedblright\ to emphasize
similarities with the big bang of the Universe) begins with very hard
initial interactions between the high-momentum partons of the colliding
nuclei. These generate the highest-momentum particles in the final state.
Afterwards, at about $t\sim 0.2$ fm/$c$, most of the soft particles in the
final state are coherently produced and form a nonequilibrium system
(sometimes called a \textquotedblleft glasma\textquotedblright\ \cite{lap06}%
)\ of high energy density. The \textquotedblleft
bottom-up\textquotedblright\ thermalization\ scenario \cite{bai01} assumes
that the initial properties of this system are determined by the QCD
saturation mechanism \footnote{%
The earlier widely used minijet initial conditions \cite{kaj87} are instead
based on an initial\ parton distribution generated by (semi-) hard
interactions between partons of the incident nuclei..}, i.e. that they are
dominated by \emph{coherent} small-$x$\ gluons of a very high density. These
gluons originate from the low-$x$ part of the nuclear wavefunctions ($x=p_{%
\text{t}}/\sqrt{s}$ where $s$ is the total energy) and carry transverse
momenta $p_{\text{t}}$ of the order of the saturation scale $Q_{\text{s}}$.
Since for RHIC collisions $Q_{\text{s}}\sim 1$ GeV $\gg \Lambda _{\text{QCD}}
$, this initial state should be amenable to weak-coupling techniques 
\footnote{%
An alternative scenario assumes that the produced matter is rather strongly
coupled \cite{shu04}. This would naturally explain the observed
thermalization rate and the surprisingly small shear viscosity (over
entropy) which dual gravity descriptions also predict.}. In the color glass
condensate (CGC) model \cite{ian03}, for example, the highly populated small-%
$x$ gluon states are treated as the soft modes of \emph{classical}
Yang-Mills fields with typically large amplitudes while the hard field modes
are represented by static sources.

More generally, as long as the gluon mode occupation numbers stay large
enough to suppress quantum effects, their evolution can be described in
terms of classical gauge fields which may be simulated fully
nonperturbatively on a spacial lattice. Over the last years, an increasing
amount of such numerical simulations was performed for the gauge groups SU$%
\left( 2\right) $ and SU$\left( 3\right) $ in one, two and three spacial
dimensions. The dynamical settings included the CGC model \cite{kra99,rom06}%
, hot-thermal-loop (HTL)\ effective theories (equivalent to a collisionless
Vlasov equation) \cite{arn03,htl,arn05} and equations of Wong-Yang-Mills
type \cite{dum05,dum07,bod07}, subject to minijet or bottom-up initial
conditions. Among the calculated observables were e.g. energy densities and
gluon multiplicity distributions.

Numerical simulations confirmed, in particular, that sufficiently
anisotropic parton momentum distributions, as typically produced in
heavy-ion collisions, can induce the onset of a very fast, collective\
isotropization and (pre-) thermalization \footnote{%
Chromo-Weibel-instability driven isotropization is reversible at the
mean-field level, i.e. it proceeds at constant entropy and thus cannot
contribute directly to equilibration. However, it shapes the parton momentum
distributions such that subsequent collision-induced thermalization becomes
much more efficient.} driven by non-Abelian filamentation instabilities \cite%
{dum05}. These Weibel-type plasma instabilities generate an initially
exponential growth of the soft-mode occupation numbers even at relatively
weak couplings \cite{mro06,mro206,plinst}. The expansion of the system can
strongly reduce this growth, however, and the non-Abelian\ self-interactions
appear to rapidly limit it (in three spacial dimensions) to at most linear
growth \cite{arn05}. This happens when the (magnetic) energy deposited in
the soft fields returns sufficiently fast to the hard fields, either by
plasmon excitation effects similar to Kolmogorov wave turbulence \cite%
{arn05,ber08} or via a rapid avalanche \cite{dum07}. Applications of the CGC
and HTL effective theories are limited to the weak-coupling and
small-amplitude regimes, however, which ensure a sufficient scale separation
between the hard parton and soft gauge-field momenta \footnote{%
This hybrid\ description allows for a simple implementation of particlelike
high-momentum fermions as well, although their impact remains limited during
the early \ thermalization stage (due to relatively small occupation
numbers). The quarks are expected to thermalize much later than the gluons 
\cite{shu92}, moreover, which we have found to be the case for scalar matter
as well.}. This technical limitation can be avoided by describing the whole
system as a classical statistical Yang-Mills ensemble with an UV lattice
cutoff to substitute for the quantum mechanical suppression\ of thermal
short-distance effects. Hard and soft modes are thus originating from the
same gauge field and treated on the same footing, which allows for
nonperturbatively large amplitudes.

The numerical simulations described in the present paper are based on the
same interpretation of the classical lattice fields. Although we focused on
the thermalization of gauge-Higgs matter by \emph{chaotic} instabilities,
which at first seems to be a rather different mechanism, there are several
commonalities with equilibration via mean-field plasma instabilities. In
fact, both are collective processes which shape the momentum distribution of
the classical gauge fields, and both lead to substantially faster gluon
equilibration rates than collisional thermalization. This becomes explicit
in relaxation times $\tau \sim \lambda ^{-1}\sim 0.5$ fm/$c$ at typical RHIC
energies. (The chaotic relaxation time tends to become even smaller for SU$%
\left( 3\right) $ gauge fields \cite{gon93}). As pointed out in Ref. \cite%
{bol00}, incidentally, relaxation times of this order imply that
fluctuations around the mean transverse momentum\ produced in nuclear
collisions are very small \footnote{%
An underlying assumption is that the lattice YMH system is globally
hyperbolic on its compact phase space, i.e. that it is an \textquotedblleft
Anosov system\textquotedblright .} (of order $10^{-2}$, i.e. at the percent
level, for typical reaction volumes (5 fm)$^{3}$ and times (0.5 fm) at
RHIC), as indeed observed in event-by-event fluctuations.

On the other hand, there are remarkable differences between chaotic and
plasma instabilities. The maximally chaotic modes are very efficient in
generating entropy \emph{directly}, for instance, while the filament
instabilities lead to a \emph{reversible} isotropization (at the mean-field
level) which just creates more efficient conditions for the subsequent
entropy production. Hence Weibel-type plasma instabilities \emph{per se}
seem hardly to affect the MLE values, although they could indirectly
contribute to our chaotic thermalization if the random initial conditions
generate a sufficiently anisotropic momentum distribution of the gauge
fields. It is tempting to speculate, then, that even under RHIC\ initial
conditions the maximally chaotic modes may lead to faster thermalization
than plasma instabilities because they generate entropy in a probably more
efficient one-step\ process. Moreover, the maximal chaotic instabilities
select at any time the most unstable direction in phase space and thus
remain optimally fast during the \emph{entire} (classical) thermalization
process, whereas the filament instabilities are damped by the gauge field's
non-Abelian self-interactions \footnote{%
After the primary instability subsides, however, it is followed by a
secondary, faster one generated by power cascades towards shorter
wavelengths which are reminiscent of Kolmogorov wave turbulence in plasmas 
\cite{ber08} (with Kolmogorov scaling exponent 3/2 as in scalar theories and
thus during the evolution of the early Universe). It would be interesting to
know whether chaotic instabilities have similar features.}.

As mentioned above, a long-term perspective of our work is to understand the
role of chaotic thermalization mechanisms during the classical evolution
phases in the aftermath of nuclear collisions. The focus of the present
paper was more limited and preparatory, however, namely to map out \emph{%
generic} chaotic thermalization properties of non-Abelian gauge systems in
the presence of scalar matter. (Classical Higgs fields were included because
of their role in the early Universe and because they can be treated on the
same footing as the gauge fields. Quarks, on the other hand, would have to
be implemented either as particles or as solutions of the Dirac equation in
the background of the classical fields, as e.g. in soliton models for
baryons \cite{lee77}.) For this reason, we have chosen random initial
conditions instead of the more specialized ones required to describe the
physical situation after a nuclear collision (or during the reheating phase
of the early Universe). Our program could naturally be extended by adapting
it to those initial conditions with a strongly anisotropic momentum
distribution which characterize the hot system created by a high-energy
nuclear collision. The chaotic properties of plasma-instability-, collision-
and cascade-driven thermalization processes could then be quantified in
terms of the corresponding maximal Lyapunov exponents, thereby relating the
chaotic thermalization and entropy production rates to the time scales of
more conventional equilibration processes. (The large difference between the
Lyapunov exponents of randomly and coherently initialized fields (cf. Secs. %
\ref{lya} and \ref{mleb})\ show that they can depend strongly on the initial
conditions.)

In addition, one may extract a more detailed picture of the chaotic
equilibration processes by following e.g. the evolution of the field modes'
momentum distributions and their anisotropy, as well as the evolution of
pressure, entropy etc. simultaneously with the Lyapunov histories. For
specific comparisons with the effective CGC and HTL dynamics, it would also
be interesting to follow the growth rate of\ the gauge field's Fourier
coefficients. This would further clarify how far chaotic evolution is
consistent with the bottom-up\ thermalization scenario, according to which
the copiously produced soft gluons (which initially carry only a small
fraction of the total energy) draw energy from the hard gluons and
thermalize very efficiently. The decay of the remaining hard gluons then
reheats the soft gluonic background until it enters the hydrodynamic
evolution stage. In fact, in reductions of the Yang-Mills dynamics to a few
degrees of freedom, chaotic thermalization was found to start (under
suitable initial conditions) among the softer modes as well \cite{bir295}.
Moreover, the\ chaotic behavior of the soft modes turns out to be driven by
the soft-hard mode coupling, which provides an efficient mechanism for
energy transfer from high frequency modes to a low-energy multiparticle
final state. (Related findings for the scattering of two classical SU$\left(
2\right) $ YM and YMH wave packets \footnote{%
The nonlinearites due to Higgs gauge-field and Higgs self-couplings, which
we found to be of relatively minor importance during the chaotic
thermalization of the YMH system, turn out to play a subordinate role in
such scattering processes as well.} were reported in Ref. \cite{hu95}.)

\section{Summary and conclusions}

\label{sum}

We have investigated chaotic instability and thermalization properties of
classical gauge and matter fields in the unbroken phase of SU$\left(
2\right) $ Yang-Mills-Higgs theory on a spacial lattice. Since equilibration
proceeds mainly through the most unstable field modes, we have focused on a
quantitative survey of the most chaotic time evolution patterns in terms of
maximal Lyapunov exponents (which measure the logarithmic distance growth
rates between initially neighboring gauge and Higgs fields).

A main goal of our investigation was to explore the impact of the
fundamental doublet of scalar matter fields on the chaotic behavior of the
gauge dynamics. Towards this end, we first confirmed and extended previous
evidence for the Yang-Mills-Higgs system to equilibrate over two drastically
different time scales: individually, the rather weakly coupled gauge and
Higgs sectors reach a preequilibrium\ phase after only a few lattice time
units, whereas their mutual equilibration is substantially delayed by the
matter fields and takes far longer than $10^{4}$ units to complete.\textbf{\ 
}Accordingly, we have generated two sets of maximal Lyapunov exponents for
initially random fields: a larger one extracted from the field separation
rates at the preequilibration stage, and a smaller one obtained from
long-time trajectories extrapolated to infinite evolution time and thus to
full equilibrium.

The first set was designed to cover a representative part of the weakly
coupled phase space and contains about 80 Lyapunov exponents. In view of its
non-negligible finite-time errors, this set was mainly used to study general
characteristics of the exponents including their energy, coupling-parameter
and lattice-size dependence. We found the signs of the whole set to be
positive (in contrast to results in the broken phase), which implies that
chaos is at least approximately universal in the symmetric phase. (Possible
exceptions may include small nonergodic niches as previously encountered in
the Yang-Mills phase space.) Yang-Mills-Higgs theories with gauge groups
containing SU$\left( 2\right) $ as a subgroup are therefore chaotic as well.
In the energy and coupling ranges where both finite-size and lattice-spacing
errors should be under control, our Lyapunov exponents on lattices with
between $10^{3}$ and $30^{3}$ sites were indeed found to be within
statistical uncertainties identical.\ In addition, we found the maximal
Lyapunov exponents extracted during the preequilibrium phase to be almost
independent of the Higgs self-coupling. This indicates that the nonlinear
interactions in the Higgs sector provide a relatively minor contribution to
the chaoticity of the system.

In order to survey the important and previously unexplored asymptotic
regions of the Yang-Mills-Higgs phase space, we have additionally followed
several field evolution trajectories over the exceptionally long periods
required to approach total equilibrium. In particular, we have investigated
the long-time behavior of eight Lyapunov histories, i.e. logarithmic
separation rates between initially neighboring fields, over 20000 lattice
units. After extrapolation to infinite evolution times they provide our best
estimates for the maximal Lyapunov exponents. All Lyapunov histories turn
out to decrease monotonically with time and saturate relatively slowly even
after the energy has attained almost complete equipartition. At larger
values of the Higgs self-coupling the Lyapunov histories vary less strongly
during time evolution while their saturation is further delayed. Moreover,
the Lyapunov histories divided by the average plaquette energy saturate
faster than the Lyapunov histories themselves. This foreshadows an early
onset of their linear scaling relation in the equilibrated system (see
below). The long-time behavior of the Lyapunov histories provides reliable
estimates for the finite-time errors of the Lyapunov exponents, furthermore,
and quantifies in particular how much they are overestimated in the
preequilibrium phase.

The physical interest in Lyapunov exponents of gauge fields originates
partly from evidence for their linear relation to the plasmon damping rates
at weak coupling. This relation relies on the empirically identified linear
dependence of the maximal Lyapunov exponents on the average magnetic
gauge-field energy. We have therefore systematically scrutinized the
accuracy and validity range of this scaling relation on the basis of our
full data set. Both during the rather long preequilibrium period, i.e.
before gauge and Higgs sectors have exchanged substantial amounts of energy,
and after full thermalization we have indeed found the Lyapunov histories to
scale within errors linearly with the average gauge-field energy. More
specifically, both during preequilibrium and after complete equilibration
our results establish the relation $\bar{\lambda}_{\text{G,H}}\simeq \bar{E}%
_{\text{p}}/6$ between the maximal Lyapunov exponents and the average
plaquette energy. This relation was previously encountered in pure
Yang-Mills theory and seems to be a rather exclusive property of non-Abelian
gauge theories. It implies that the Lyapunov exponents for any given \emph{%
gauge} field energy, extracted either from the gauge or Higgs field
separation rates, remain within errors independent of the presence of the
Higgs field. This lends additional credence to the suggested equality
between the Lyapunov exponents and twice the static plasmon damping rates of
quantum Yang-Mills-Higgs theory at high temperatures and weak coupling. In
addition, we have established that both during the preequilibrium stage and
after full thermalization the maximal Lyapunov exponents also scale linearly
with the total energy. This is in contrast to previous expectations and
turns out to be a consequence of the proportionality between gauge and total
energy during both phases.

We have furthermore studied how interactions with the Higgs field affect the
Savvidy instability of constant non-Abelian magnetic fields. As a benchmark,
we have first obtained the maximal Lyapunov exponent for an initially
constant magnetic field in pure Yang-Mills theory and found it about twice
as large as that of a randomly initialized field under otherwise equal
conditions. We have then computed the analogous Lyapunov exponent in
Yang-Mills-Higgs theory for initially homogeneous magnetic and Higgs fields
at the same total energy and found it still to be about 2 times larger than
for randomized fields. Hence the additional matter fields seem neither to
dampen nor to enhance the Savvidy instability relative to that of random
fields of the same energy. Depending on the initial energy distribution
between gauge and Higgs fields, however, the presence of matter fields can
have a strong impact on the absolute magnitude of the magnetic field's
instability. In order to demonstrate this, we have provided an example in
which the matter fields approximately double the maximal Lyapunov exponent
of the gauge field.

In all simulations described above, we found the Lyapunov histories to be
within errors of at most a few percent independent of the underlying
distance measure and of whether they were obtained in the gauge or Higgs
field's phase space. This independence turns out to hold both during the
preequilibration phase and towards full equilibrium, and both for initially
homogeneous and random fields. Hence it confirms the general expectation
that the divergence rates of the most unstable modes should be equally
measurable by any reasonable metric in field space. The behavior of the
Lyapunov histories depends somewhat on the stage of the thermalization
process, however. Before gauge and Higgs fields have exchanged substantial
amounts of energy, the divergence rate in the gauge sector turns out to be
slightly\ larger than in the Higgs sector, while it becomes marginally
smaller during later phases of equilibration. A previous estimate of an
about 15\% smaller logarithmic Higgs field\ separation rate, based on a
single trajectory, was therefore probably contaminated by numerical
uncertainties.

Our above results strengthen the evidence for the gauge dynamics to provide
the main source of chaotic instability in the Yang-Mills-Higgs system. The
matter fields, in contrast, seem to play a subordinate role (similar to the
quark fields shortly after a nuclear collision). The evidence includes the
fast preequilibration of the gauge sector, the observation that the Higgs
sector seems to have little impact on the ratio between the constant
magnetic and random field Lyapunov exponents, the linear dependence of the
maximal Lyapunov exponents on the average gauge-field energy alone, the
finding that this scaling behavior sets in way before the gauge fields have
full access to the energy stored in the Higgs sector, and in particular the
fact that the values of the maximal Lyapunov exponents for a given plaquette
energy turn out to be within errors identical to those in pure Yang-Mills\
theory. Moreover, the nonlinear Higgs dynamics seems not to contribute
substantially to the maximally unstable mode, at least not in the range of
relatively weak couplings where lattice artefacts are under control. Indeed,
at the preequilibration stage the Lyapunov histories turn out to be
practically independent of the Higgs self-coupling, and even afterwards the
coupling dependence remains moderate.

Nevertheless, we found the scalar\ matter fields to have a major impact on
the thermalization of the gauge system. As we have shown, their presence can
strongly enhance the absolute divergence rate between neighboring gauge
fields, both homogeneous and random. Moreover, the maximally chaotic mode
has an almost immediate effect on the Higgs sector, as witnessed by the fact
that its separation rate can be monitored equally well by following the
distance evolution between neighboring Higgs fields. Finally, the presence
of the Higgs fields massively prolongs the equilibration of the system as a
whole, at least at weak coupling. This qualitative effect may be robust
enough to prevail in the case of fermionic matter, and hence be relevant for
understanding the\ equilibration properties of the highly excited
quark-gluon matter produced in ultrarelativistic nuclear collisions.

\begin{acknowledgments}
R.F. would like to express his particular gratitude to Gast\~{a}o Krein for
their discussions and for support and encouragement. Financial support from
the Funda\c{c}\~{a}o de Amparo \`{a} Pesquisa do Estado de S\~{a}o Paulo
(FAPESP) under Grant No. 04/13405-0 and from the Deutsche
Forschungsgemeinschaft (DFG) is also acknowledged.
\end{acknowledgments}

\newpage

\begin{thebibliography}{99}
\bibitem{hei04} U. Heinz, arXiv:hep-ph/0407360; P. Jacobs and X.-N. Wang,
Prog. Part. Nucl. Phys. \textbf{54}, 443 (2005).

\bibitem{bas06} B.A. Bassett, S. Tsujikawa and D. Wands, Rev. Mod. Phys. 
\textbf{78}, 537 (2006).

\bibitem{lin80} A.D. Linde, Phys. Lett. \textbf{B 96}, 289 (1980).

\bibitem{sel93} A.V. Selikhov and M. Gyulassy, Phys. Lett. \textbf{B 316},
373 (1993).

\bibitem{bra90} E. Braaten and R.D. Pisarski, Phys. Rev. D \textbf{42}, 2156
(1990).

\bibitem{pis08} R.D. Pisarski, arXiv:0810.4585.

\bibitem{mul92} B.\ M\"{u}ller and A.\ Trayanov, Phys.\ Rev.\ Lett. \textbf{%
68}, 3387 (1992).

\bibitem{bir294} T.S. Bir\'{o}, S.G. Matinyan, and B. M\"{u}ller, \textit{%
Chaos and Gauge Field Theory}, World Scientific, Singapore, 1994; T.S. Bir%
\'{o}, \'{A}. F\"{u}l\"{o}p, C. Gong, S.G. Matinyan, B. M\"{u}ller and A.\
Trayanov, Lecture Notes in Physics \textbf{494}, 164 (1997).

\bibitem{bir95} T. S. Bir\'{o}, C. Gong and B. M\"{u}ller, Phys. Rev. D 
\textbf{52}, 1260 (1995).

\bibitem{bla02} J.P. Blaizot and E. Iancu, Phys. Rep. \textbf{359}, 355\
(2002).

\bibitem{pro97} Proceedings of the Workshop on Strong and Electroweak
Matter, Eger, Hungary, 1997, edited by F. Csikor and Z. Fodor (World
Scientific, Singapore, 1998); M. Dine and A. Kusenko, Rev. Mod. Phys. 
\textbf{76}, 1 (2003).

\bibitem{hin95} M.B. Hindmarsh and T.W.B. Kibble, Rept. Prog. Phys. \textbf{%
58}, 477 (1995); I. Tkachev, S. Khlebnikov, L. Kofman and A.D. Linde, Phys.
Lett. \textbf{B 440}, 262 (1998).

\bibitem{arm08} N. Armesto et al., J. Phys. G \textbf{35}, 054001 (2008).

\bibitem{mro06} S. Mr\'{o}wczy\'{n}ski, Phys. Lett. \textbf{B 393}, 26
(1997); Acta Phys. Pol. B \textbf{39}, 1665 (2008).

\bibitem{amb91} J. Ambjorn, T. Askgaard, H. Porter and M.E. Shaposhnikov,
Nucl. Phys. \textbf{B 353}, 346 (1991).

\bibitem{for07} H. Forkel, Phys. Rev. D \textbf{73}, 105002\ (2006); Int. J.
Mod. Phys. E \textbf{16}, 2789 (2007).

\bibitem{far05} R. Fariello, H. Forkel and G. Krein, Phys. Rev. D \textbf{72}%
, 105015\ (2005); H. Forkel, arXiv:0810.2098, Int. J. Mod. Phys. D, to
appear.

\bibitem{kog75} J. Kogut and L. Susskind, Phys. Rev. D \textbf{11}, 395
(1975).

\bibitem{chi85} S.A. Chin, O.S. van Roosmalen, E.A. Umland and S.E. Koonin,
Phys. Rev. D \textbf{31}, 3201 (1985).

\bibitem{bir94} T.S.\ Bir\'{o}, C.\ Gong, B.\ M\"{u}ller and A.\ Trayanov,
Int.\ J.\ Mod.\ Phys. C \textbf{5}, 113 (1994).

\bibitem{ben80} G. Benettin, L. Galgani, A. Giorgilli and J.-M. Strelcyn,
Meccanica \textbf{15}, 21 (1980); M. D. Hartl, Phys. Rev. D \textbf{67},
024005 (2003).

\bibitem{gon94} C. Gong, Phys. Rev. D \textbf{49}, 2642 (1994). For more
recent simulations on somewhat larger lattices see Ref. \cite{bol00} for the
real Lyapunov spectrum and Ref. \cite{ful01} for the complex Lyapunov
spectrum (and the Kolmogorov-Sinai entropy) from the eigenvalues of the
monodromy matrix.

\bibitem{bol00} J. Bolte, B. M\"{u}ller and A. Sch\"{a}fer, Phys. Rev. D 
\textbf{61}, 054506 (2000).

\bibitem{ful01} \'{A}. F\"{u}l\"{o}p and T.S. Bir\'{o}, Phys. Rev. C \textbf{%
64}, 064902 (2001)

\bibitem{mue96} B. M\"{u}ller, Duke University Report No. DUKE-TH-96-118,
chao-dyn/9607001 (unpublished).

\bibitem{hei97} U.\ Heinz, C.R.\ Hu, S.\ Leupold, S.G.\ Matinyan and B.\ M%
\"{u}ller, Phys.\ Rev. D \textbf{55}, 2464 (1997).

\bibitem{bir96} T.\ S.\ Bir\'{o} and M.H.\ Thoma, Phys.\ Rev. D \textbf{54},
3465 (1996).

\bibitem{amb92} J. Ambjorn, T. Aksgaard, H. Porter and M.E. Shaposhnikov,
Nucl. Phys. \textbf{B 353}, 346 (1991).

\bibitem{nie96} H.B. Nielsen, H.H. Rugh and S.E. Rugh,
arXiv:chao-dyn/9605013; arXiv:hep-th/9611128.

\bibitem{zak94} J. Zakrzewski, Phys. Lett. \textbf{B 327}, 67 (1994).

\bibitem{dah90} P. Dahlqvist and G. Russberg, Phys. Rev. Lett. \textbf{65},
2837 (1990).

\bibitem{mat81} S.G. Matinyan, G.K. Savvidi and N.G. Ter-Arutyunyan-Savvidi,
JETP Lett. \textbf{34}, 590\ (1981).

\bibitem{ber85} G.P. Berman, Yu.I. Man'kov and A.F. Sadreev, Sov. Phys. JETP 
\textbf{61}, 415\ (1985).

\bibitem{mat97} S.G. Matinyan and B. M\"{u}ller, Phys. Rev. Lett. \textbf{78}%
, 2515 (1997).

\bibitem{gon93} C. Gong, Phys. Lett. \textbf{B 298}, 257 (1993).

\bibitem{kra96} A. Krasnitz, Nucl. Phys. \textbf{B 455}, 320 (1995).

\bibitem{sav77} G. K. Savvidy, Phys. Lett. \textbf{B 71}, 133 (1977).

\bibitem{tra92} A.\ Trayanov and B.\ M\"{u}ller, AIP Conference Proceedings 
\textbf{260}, 280 (1992).

\bibitem{chaoconf} See e.g. H.B. Nielsen and P. Olesen, Nucl. Phys. \textbf{%
B 160}, 380\ (1979); G.Z. Baseyan, S.G. Matinyan, and G.K. Savvidy, Sov.
Phys. JETP \textbf{53}, 247 (1979) [Zh. Eksp. Teor. Fiz. \textbf{29}, 641
(1979)]; J. Ambjorn and P. Olesen, Nucl. Phys. \textbf{B 170}, 60\ (1980);
S.G. Matinyan, G.K. Savvidy, and N.G. Ter-Arutyunyan-Savvidy, ibid, \textbf{%
80}, 830 (1981); Sov. Phys. JETP \textbf{53}, 421 (1981) [Zh. Eksp. Teor.
Fiz. \textbf{80}, 830 (1981)]; P. Olesen, Nucl. Phys. \textbf{B 200}, 381
(1982); G.K. Savvidy, Nucl. Phys. \textbf{B 246}, 302 (1984); W.H. Steeb,
J.A. Louw, and C.M. Villet, Phys. Rev. D 33, 1174 (1986); A. Giansanti and
P.D. Simic, Phys. Rev. D \textbf{38}, 1352 (1988); T. Biro, M. Feurstein and
H. Markum, Heavy Ion Phys. \textbf{7}, 235 (1998).

\bibitem{inf} A.D. Linde,\emph{\ Particle Physics and Inflationary Cosmology}%
, Harwood, Chur, Switzerland (1990); A.R. Liddle and D.H. Lyth, \emph{%
Cosmological inflation and large-scale structure}, Cambridge University
Press, Cambridge (2000); J. Gonzalo, \emph{Inflationary cosmology revisited}%
, World Scientific, Singapore (2005); M. Lemoine, J. Martin and P. Peter
(eds.), \emph{Inflationary Cosmology}, Lect. Notes in Physics 0738,
Springer, New York (2008).

\bibitem{kof94} L. Kofman, A.D. Linde and A.A. Starobinsky, Phys. Rev. Lett. 
\textbf{73}, 3195 (1994); \textbf{76}, 1011 (1996); Phys. Rev. D \textbf{56}%
, 3258 (1997); Y. Shtanov, J.H. Traschen and R.H. Brandenberger, Phys. Rev.
D \textbf{51}, 5438 (1995); G.N. Felder, J. Garcia-Bellido, P.B. Greene, L.
Kofman, A.D. Linde and I. Tkachev, Phys. Rev. Lett. \textbf{87}, 011601
(2001).

\bibitem{fel01} G. Felder, J. Garc\'{\i}a-Bellido, P. Greene, L. Kofman,
A.D. Linde and I. Tkachev, Phys. Rev. Lett. \textbf{87}, 011601 (2001); G.
Felder, L. Kofman and A.D. Linde, Phys. Rev. D \textbf{64}, 123517 (2001).

\bibitem{mic04} R. Micha and I.I. Tkachev, Phys. Rev. D \textbf{70}, 043538
(2004).

\bibitem{pod06} D. Podolsky, G.N. Felder, L. Kofman and M. Peloso, Phys.
Rev. D \textbf{73}, 023501 (2006).

\bibitem{ber04} J. Berges, Sz. Bors\'{a}ny and C. Wetterich, Phys. Rev.
Lett. \textbf{93}, 142002 (2004).

\bibitem{dia08} A. D\'{\i}az-Gil, J. Garc\'{\i}a-Bellido, M. Garc\'{\i}a-P%
\'{e}rez and A. Gonz\'{a}lez-Arroyo, Phys. Rev. Lett. \textbf{100}, 241301
(2008).

\bibitem{preh} J. Garc\'{\i}a-Bellido, M. Garc\'{\i}a P\'{e}rez and A. Gonz%
\'{a}lez-Arroyo, Phys. Rev. D \textbf{69}, 023504 (2004); A. D\'{\i}az-Gil,
J. Garc\'{\i}a-Bellido, M. Garc\'{\i}a P\'{e}rez and A. Gonz\'{a}lez-Arroyo,
Phys. Rev. Lett. \textbf{100}, 241301 (2008); A. Rajantie, P.M. Saffin and
E.J. Copeland, Phys. Rev. D \textbf{63}, 123512 (2001); E.J. Copeland, D.
Lyth, A. Rajantie and M. Trodden, Phys. Rev. D \textbf{64}, 043506 (2001);
A. Tranberg and J. Smit, JHEP \textbf{08}, 012 (2006); A. Tranberg, J. Smit
and M. Hindmarsh, JHEP \textbf{01}, 034 (2007).

\bibitem{esb} W.H. Tang and J. Smit, Nucl. Phys. \textbf{B 510} (1998) 401;
G.D. Moore and N. Turok, Phys. Rev. D \textbf{55} (1997) 6538; G.D. Moore
and K. Rummukainen, Phys. Rev. D \textbf{63} (2001) 045002; M. Hindmarsh and
A. Rajantie, Phys. Rev. D \textbf{64} (2001) 065016.

\bibitem{kuz85} V.A. Kuzmin, V.A. Rubakov and M.E. Shaposhnikov, Phys. Lett. 
\textbf{B 155}, 36 (1985).

\bibitem{tan96} W.H. Tang and J. Smit, Nucl. Phys. \textbf{B 482}, 265
(1996).

\bibitem{amb97} J. Ambj\o rn and A. Krasnitz, Nucl. Phys. \textbf{B 506},
387 (1997); G.D. Moore and N. Turok, Phys. Rev. D \textbf{56}, 6533 (1997).

\bibitem{moo99} G.D. Moore, Phys. Rev. D \textbf{59}, 014503 (1998); D. B%
\"{o}deker, G.D. Moore and K. Rummukainen, Phys. Rev. D \textbf{61}, 056003
(2000); B.-J. Nauta and A. Arrizabalaga, Nucl. Phys. \textbf{B 635}, 255
(2002).

\bibitem{gar99} J. Garcia-Bellido, D.Y. Grigoriev, A. Kusenko and M.E.
Shaposhnikov, Phys. Rev. D \textbf{60}, 123504 (1999).

\bibitem{gar04} J. Garcia-Bellido, M. Garcia Perez and A. Gonzalez-Arroyo,
Phys. Rev. D \textbf{69}, 023504 (2004).

\bibitem{moo01} G.D. Moore, JHEP \textbf{11}, 021 (2001).

\bibitem{tra03} A. Tranberg and J. Smit, JHEP \textbf{11}, 016 (2003).

\bibitem{sku03} J.-I. Skullerud, J. Smit and A. Tranberg, JHEP \textbf{08},
045 (2003).

\bibitem{cas} X.-N. Wang, Phys. Rep. \textbf{280}, 287 (1997);\ K. Geiger,
Phys. Rep. \textbf{258}, 237 (1995); B. Zhang, Comput. Phys. Commun. \textbf{%
104}, 70 (1997).

\bibitem{shu92} E. Shuryak, Phys. Rev. Lett. \textbf{68}, 3270\ (1992).

\bibitem{mro206} P. Arnold, Int. J. Mod. Phys. E \textbf{16}, 2555 (2007);
S. Mr\'{o}wczy\'{n}ski, PoS (CPOD2006) 042, (2006).

\bibitem{rhic} I. Arsene \emph{et al.} (BRAHMS Collaboration), Nucl. Phys. 
\textbf{A 757}, 1 (2005); K. Adcox \emph{et al.} (PHENIX Collaboration),
Nucl. Phys. \textbf{A 757}, 184 (2005); B. B. Back \emph{et al.} (PHOBOS
Collaboration), Nucl. Phys. \textbf{A 757}, 28 (2005); J. Adams \emph{et al.}
(STAR Collaboration), Nucl. Phys. A \textbf{757}, 102 (2005).

\bibitem{arn205} P. Arnold, J. Lenaghan, G.D. Moore and L.G. Yaffe, Phys.
Rev. Lett. \textbf{94}, 072302 (2005).

\bibitem{arn03} P. Arnold, J. Lenaghan and G.D. Moore, JHEP \textbf{08}, 002
(2003).

\bibitem{lap06} T. Lappi and L. McLerran, Nucl. Phys. \textbf{A 772}, 200
(2006).

\bibitem{bai01} R. Baier, A.H. Mueller, D. Schiff and D.T. Son, Phys. Lett. 
\textbf{B 502}, 51 (2001); Phys. Lett. \textbf{B 539}, 46 (2002).

\bibitem{kaj87} K. Kajantie, P.V. Landshoff and J. Lindfors, Phys. Rev.
Lett. \textbf{59}, 2527 (1987);\ K.J. Eskola, K. Kajantie and J. Lindfors,
Nucl. Phys. \textbf{B 323}, 37 (1989); J.-P. Blaizot and A.H. Mueller, Nucl.
Phys. \textbf{B 289}, 847 (1987).

\bibitem{shu04} E. Shuryak, J. Phys. G \textbf{30}, S1221 (2004).

\bibitem{ian03} E. Iancu and R. Venugopalan, in \emph{Quark gluon plasma 3},
ed. by R.C. Hwa and X.N. Wang, World Scientific, Singapore (2003).

\bibitem{kra99} A. Krasnitz and R. Venugopalan, Nucl. Phys. \textbf{B 557},
237 (1999); Phys. Rev. Lett. \textbf{84}, 4309 (2000); ibid. \textbf{86},
1717 (2001); ibid. \textbf{87}, 192302 (2001).

\bibitem{rom06} P. Romantschke and R. Venugopalan, Phys. Rev. Lett. \textbf{%
96}, 062302 (2006);\ Phys. Rev. D \textbf{74}, 045011 (2006).

\bibitem{htl} S. Mr\'{o}wczy\'{n}ski, Phys. Lett. \textbf{B 214}, 587
(1988); S. Mr\'{o}wczy\'{n}ski, A. Rebhan, and M. Strickland, Phys. Rev. D 
\textbf{70}, 025004 (2004); P. Romatschke and A. Rebhan, Phys. Rev. Lett. 
\textbf{97}, 252301 (2006); B. Schenke and M. Strickland, Phys. Rev. D 
\textbf{74}, 065004 (2006); B. Schenke, M. Strickland, C. Greiner, and M. H.
Thoma, Phys. Rev. D \textbf{73}, 125004 (2006).

\bibitem{arn05} P. Romatschke and M. Strickland, Phys. Rev. D \textbf{68},
036004 (2003); \textbf{70}, 116006 (2004); P. Arnold, G.D. Moore and L.G.
Yaffe, Phys. Rev. D \textbf{72}, 054003 (2005);\ A. Rebhan, P. Romantschke
and M. Strickland, JHEP \textbf{09}, 041 (2005);\ P. Arnold and G.D. Moore,
Phys. Rev. D \textbf{73}, 025006 (2006); \textbf{73}, 025013 (2006).

\bibitem{dum05} A. Dumitru and Y. Nara, Phys. Lett. \textbf{B 621}, 89
(2005).

\bibitem{dum07} A. Dumitru, Y. Nara and M. Strickland, Phys. Rev. D \textbf{%
75}, 025016 (2007).

\bibitem{bod07} D. B\"{o}deker and K. Rummukainen, JHEP \textbf{07}, 022
(2007).

\bibitem{plinst} A. Rebhan, P. Romatschke, and M. Strickland, Phys. Rev.
Lett. \textbf{94}, 102303 (2005); A. Rebhan, P. Romatschke and M.
Strickland, JHEP \textbf{09}, 041 (2005); P. Arnold and G.D. Moore, Phys.
Rev. D \textbf{76}, 045009 (2007); J. Berges, D. Gelfand, S. Scheffler and
D. Sexty, Phys. Lett. \textbf{B 677}, 210 (2009).

\bibitem{ber08} J. Berges, S. Scheffler and D. Sexty, arXiv:0811.4293

\bibitem{lee77} R. Friedberg and T.D. Lee, Phys. Rev. D \textbf{15}, 1694
(1977); Phys. Rev. D \textbf{16}, 1096 (1977); L. Bayer, H. Forkel and W.
Weise, Z. Phys. A \textbf{324}, 365 (1986).

\bibitem{bir295} T.S. Bir\'{o}, S.G.\ Matinyan and B. M\"{u}ller, Phys.
Lett. \textbf{B 362}, 29 (1995).

\bibitem{hu95} C.R. Hu, S.G. Matinyan, B. M\"{u}ller, A. Trayanov, T.M.
Gould, S.D.H. Hsu and E.R. Poppitz, Phys. Rev. D \textbf{52}, 2402 (1995);
C.R. Hu, S.G. Matinyan, B. M\"{u}ller, Phys. Rev. D \textbf{54}, 2175
(1996). 
\end{thebibliography}
\end{document}